\documentclass[useAMS,usenatbib]{mn2e}
\usepackage{times,epsfig} 
\usepackage{natbib}

\newcommand{\aj}{AJ}
\newcommand{\apj}{ApJ}
\newcommand{\apjl}{ApJ}
\newcommand{\apjs}{ApJS}
\newcommand{\aap}{A\&A}

\newcommand{\mnras}{MNRAS}
\newcommand{\nat}{Nature}

\newcommand{\prd}{Phys. Rev. D}

\title[Galaxy clusters in dark energy cosmologies]{
Hydrodynamical simulations of galaxy clusters in dark energy cosmologies: I. general properties}

\author[]
{\parbox[]{6.in} {C. De Boni$^1$\thanks{E-mail: cristiano.deboni@unibo.it}, K. Dolag$^2$, S. Ettori$^{3,4}$, L. Moscardini$^{1,3,4}$, V. Pettorino$^5$, \\ C. Baccigalupi$^{5,6}$  \\
\footnotesize
$^1$ Dipartimento di Astronomia, Universit\`a di Bologna, via Ranzani 1,
  I-40127 Bologna, Italy \\
$^2$ Max-Planck-Institut f\"ur Astrophysik, Karl-Scwarzschild Strasse 1, 
  D-85741 Garching bei M\"unchen, Germany \\  
$^3$ INAF, Osservatorio Astronomico di Bologna, via Ranzani 1,
  I-40127 Bologna, Italy  \\
$^4$ INFN, Sezione di Bologna, viale Berti Pichat 6/2,
  I-40127 Bologna, Italy  \\
$^5$ SISSA/ISAS, Via Bonomea 265, I-34136 Trieste, Italy \\ 
$^6$ INFN, Sezione di Trieste, via Valerio 2,
  I-34127 Trieste, Italy \\
}}                                            
\date{}

\begin{document}

\maketitle

\begin{abstract}
We investigate the influence of dark energy on structure formation, within five different cosmological models, namely a concordance $\Lambda$CDM
model, two models with dynamical dark energy, viewed as a quintessence scalar 
field (using a RP and a SUGRA potential form) and two extended quintessence models (EQp and EQn) where the quintessence scalar field interacts non-minimally with gravity
(scalar-tensor theories). We adopted for all models the
normalization of the matter power spectrum $\sigma_{8}$ to
match the CMB data. In the models with dynamical dark energy and quintessence, we describe the equation of state with $w_0\approx-0.9$, still within the range allowed by observations. For each model, we have performed hydrodynamical simulations in a cosmological box of $(300 \ {\rm{Mpc}} \ h^{-1})^{3}$ including baryons and allowing for cooling and star formation. The contemporary presence of evolving dark energy and baryon physics allows us to investigate the interplay between the different background cosmology and the evolution of the luminous
matter. Since cluster baryon fraction can be used to constrain other cosmological parameters such as $\Omega_{m}$, we also analyse how dark energy influences the baryon content of galaxy clusters.
We find that, in models with dynamical dark energy, the evolving cosmological background leads to different star formation rates and different formation histories of galaxy clusters, but the baryon physics is not affected in a relevant way. We investigate several proxies for the cluster mass function based on X-ray observables like temperature, luminosity, $M_{gas}$, and $Y_{X}$. We conclude that the X-ray temperature and $M_{gas}$ functions are better diagnostic to disentangle the growth of structures among different dark energy models. We also evaluate the cosmological volumes needed to distinguish 
the dark energy models here investigated using the cluster number counts (in terms of the mass function and the X-ray luminosity and temperature functions).
Relaxed, massive clusters, when studied in regions sufficiently far from from the centre, are built up in a very similar way despite the different dark energy models here considered.
We confirm that the overall baryon fraction is almost independent of the dark energy models at a few percent level. The same is true for the gas fraction.
This evidence reinforces the use of galaxy clusters as cosmological probe of the matter and energy content of the Universe.
\end{abstract}


\section{Introduction} 
 
Over the last decade great observational evidence \citep{1998AJ....116.1009R,1999ApJ...517..565P,2011ApJS..192...14J,2009ApJ...692.1060V} has shown that at the present time the Universe is expanding at an accelerated rate. This fact can be attributed to a component with negative pressure, which is usually referred to as dark energy, that today accounts for about 3/4 of the entire energy budget of the Universe. The simplest form of dark energy is a cosmological constant term $\Lambda$ in Einstein's equation, within the so-called $\Lambda$CDM cosmologies. Though in good agreement with observations, a cosmological constant is theoretically difficult to understand in view of the fine-tuning and coincidence problems. A valid alternative consists in a dynamical dark energy contribution that changes in time and space, often associated to a scalar field (the `cosmon' or `quintessence') evolving in a suitable potential \citep{1988NuPhB.302..645W, 1988PhRvD..37.3406R}. Dynamical dark energy allows for appealing scenarios in which the scalar field is the mediator of a fifth
force, either within scalar-tensor theories or in interacting scenarios \citep[and references therein]{1995A&A...301..321W, 2000PhRvD..62d3511A, 2000PhRvL..85.2236B,2008PhRvD..77j3003P, 2008PhLB..663..160M}. 
In view of future observations, it is of fundamental interest to investigate whether dark energy leaves some imprints in structure formation, giving a practical way to distinguish among different cosmologies, as recently investigated in \cite{2007PhRvD..76f4004H}, \cite{2010MNRAS.403.1684B}, \cite{2010ApJ...712L.179Z}, \cite{2011MNRAS.411.1077B}, \cite{2011MNRAS.412L...1B} and \cite{2010PhRvD..81f3525W}.

In this paper, we study the general properties of galaxy clusters in different dark energy cosmologies. Galaxy clusters are the largest virialized objects in the Universe and are considered to be a fair sample of the overall matter distribution of the Universe itself. They contain a large amount of gas in the form of diffused ionized plasma known as intracluster medium (ICM), which emits in the X-ray band. The X-ray properties of galaxy clusters such as luminosity and temperature trace the total mass of the cluster itself, and hence can be used to study global properties of these objects. A lot of observational work (Chandra, XMM-Newton) has been made in recent years, and future missions ({\it{e.g.}} IXO, eROSITA, WFXT) are under study to improve the characterization of these objects in the X-rays. The properties of galaxy clusters, in particular their mass, can be investigated also in the optical region of the spectrum through gravitational lensing, which gives independent estimates from X-rays.

In this paper, we analyse the properties of simulated galaxy clusters in cosmologies with different dark energy models. We follow the formation and evolution of structures in hydrodynamical simulations of a cosmological box of size $(300 \ {\rm{Mpc}} \ h^{-1})^{3}$ for five different cosmologies. These cosmologies are in general characterized by the presence of a dynamical dark energy component, {\it{i.e.}} a dark energy component with density and equation of state evolving with time, and they will be introduced and discussed in detail in Sect. \ref{models}. After presenting the selection and composition of the sample we extract from the simulations, we will study the mass function at different redshifts. Different dark energy models have a different CMB normalization of the spectrum of the perturbations and a different growth factor, both things affecting the mass function of galaxy clusters. 
Then we will study the X-ray luminosity and temperature functions, because both quantities are a proxy for the cluster mass and can therefore be used to search for dark energy imprints. In this case the advantage is that these quantities are somehow directly observable with space facilities, while evaluating the mass of a galaxy cluster requires the assumption of a model or physical hypotheses (such as hydrodynamical equilibrium) that can introduce systematics (for example if the cluster is not in a relaxed state), as shown in \cite{2006MNRAS.369.2013R}. We will also examine the X-ray luminosity-temperature ($L-T$) relation of our sample, in order to check whether there is any clear discrepancy between the properties of the simulated objects in comparison with the observed ones \citep{1999MNRAS.305..631A,2007ApJ...668..772M,2009A&A...498..361P}. After the analysis of the global properties of the simulated sample, we will concentrate on the internal properties of the single clusters and study the relative distribution of the different mass components. In particular we will check the dark energy dependence of the baryon fraction $f_{bar}$. The study of the baryon fraction in galaxy clusters, either in the form of ICM or of stars in galaxies, is useful to understand the formation history and the properties of these objects. Combined with other independent information, such as for example the value of the baryonic density $\Omega_{b}$ derived from primordial nucleosynthesis, $f_{bar}$ can be used to derive the value of the matter density of the Universe, $\Omega_{m}=\Omega_{b} f_{bar}$.

The paper is organized as follows. In Sect. \ref{models} the different cosmological models will be introduced; in Sect. \ref{simulations} the simulation set will be discussed and in Sect. \ref{mass} the study of the mass functions of the selected sample will be analysed. Sect. \ref{LT} is centred on the analysis of the X-ray $L-T$ relation, while in Sect. \ref{functions} the X-ray observables functions will be studied. In Sect. \ref{bias} we will describe the analysis and the results of the study of the cluster baryon fraction, while conclusions will be drawn in Sect. \ref{conclusions}.

\section{The cosmological models} \label{models}

In a homogeneous and isotropic universe, the cosmological expansion depends on the various energy components, as described by the Friedmann equation:
\begin{eqnarray}
\label{friedmann}
H^2 &=&H_{0}^2\left[\sum_{i}\Omega_{0i}\exp\left(-3\int_{a_{0}}^{a}\frac{1+w_{i}(a')}{a'}da'\right)
  \right. \nonumber \\
&& + \left. \left(\frac{a_{0}}{a}\right)^2(1-\sum_{i}\Omega_{0i})\right] \ .
\end{eqnarray} 

\noindent Here, the Hubble parameter is defined as $H \equiv \dot{a}/a$
(where the dot denotes the derivative with respect to the cosmic time
$t$), $\Omega_{0i} \equiv \rho_{0i}/\rho_{0c}$ is the current density
parameter of the $i$-th component of the universe, $\rho_{0c} \equiv 3H_{0}^2/8 \pi G$ is the
critical density at the present time, $w_{i}$ is the equation of state parameter of
the $i$-th component ($w_{i} \equiv p_{i} / \rho_{i}$) and the sum is taken
over all components. We have used natural units and set $c=1$.
The first term in equation (\ref{friedmann}) includes densities associated to
each constituent of the universe while the second term accounts for
any possible deviation from flat geometry. We can assume that the
universe is constituted of three different components: matter [baryons
and cold dark matter (CDM)], for which $w_{m}=0$; radiation (photons
plus relativistic matter) with $w_{r}=1/3$, whose contribution is nowadays negligible; dynamical dark energy (DE), which, in the simplest case, behaves as a fluid with negative time
dependent $w_\mathrm{DE}$ and provides the present accelerated expansion of the
universe. We consider flat cosmological models for which
$\sum_{i}\Omega_{0i}=1$, so the curvature term in the Friedmann
equation will be equal to zero. With this in mind, and introducing
$\kappa \equiv 8 \pi G$, we can express equation (\ref{friedmann}) as

\begin{equation}
\label{friedmann_density}
H^2 =
\frac{\kappa}{3}\sum_{i}\rho_{0i}\exp\left(-3\int_{a_{0}}^{a}\frac{1+w_{i}(a')}{a'}da'\right)
\ .
\end{equation}

For our analysis we have considered three possible sets of cosmological models. The first is the standard $\Lambda$CDM model, that we use as a reference model, where dark energy is represented by the cosmological constant. This model is in agreement with present observations, though theoretically it is
intrinsically affected by fine-tuning and coincidence problems. Alternatively, dark energy could be a dynamical component, seen as a scalar field rolling down a potential \citep{1988NuPhB.302..645W,1988PhRvD..37.3406R}. If the scalar field is
minimally coupled to gravity, this class of scenarios is still affected by
fine-tuning and coincidence problems, as much as in the standard $\Lambda$CDM
model. However it is interesting, for our analysis, to consider such
dynamical cases, where a time varying equation of state is present. Numerical simulations of quintessential cold dark matter have been presented, for example, in \cite{2010MNRAS.401.2181J}. More interestingly, the dynamical scalar field could be coupled to other species, as addressed in \cite{1995A&A...301..321W}, \cite{2000PhRvD..62d3511A} and \cite{2008PhRvD..77j3003P}. We limit ourselves to the case in which the coupling involves universally all species, as it happens in scalar-tensor theories \citep{2000PhRvL..85.2236B}. The latter have been also investigated within $F(R)$ theories in
\cite{2009PhRvD..79h3518S} and \cite{2008PhRvD..78l3524O}. $N$-body simulations with a coupling to gravity (extended quintessence) have been studied, for example, in \cite{2011ApJ...728..109L}. Note that hydro-simulations including a coupling to dark matter have been presented in \cite{2010MNRAS.403.1684B}. $N$-body simulations for coupled dark energy have been investigated in
\cite{2004PhRvD..69l3516M} and \cite{ 2010ApJ...712L.179Z}. In
\cite{ 2009PhRvD..80d4027L} and \cite{2010ApJ...712L.179Z} the effect of scalar
field perturbations was also taken into account. Further fifth force
couplings have been simulated in \cite{2009PhRvD..80d4027L}. The impact of early dark energy in structure formation has been considered in \cite{2009MNRAS.394.1559G}. Our analysis differs from
previous works because of different dynamics in the dark energy models considered and of the contemporary inclusion of baryonic physics in the framework of evolving dark energy scenarios.

We will now specify in more detail the cosmologies here considered and our choice of the parameters.

\subsection{$\Lambda$CDM} \label{LCDM}

The first model considered is the concordance $\Lambda$CDM model, with
the values of the cosmological parameters taken from WMAP3 (see Subsect. \ref{choice}). This model is characterized by the presence of a dark energy component given by a cosmological constant $\Lambda$, with a constant $w_{\Lambda}=-1$, so that equation (\ref{friedmann_density}) can be written as

\begin{equation}
H^2=\frac{\kappa}{3}\left[\rho_{0m}\left(\frac{a_{0}}{a}\right)^3+\rho_{0r}\left(\frac{a_{0}}{a}\right)^4+\rho_{0\Lambda}\right] \ ,
\end{equation}

\noindent with $m$, $r$, and $\Lambda$ denoting the matter, radiation and cosmological constant components respectively.

\subsection{Minimally coupled quintessence} \label{OQ}

The second case that we consider here is that of a dynamical dark energy, given by a quintessence scalar field $\phi$ with an equation of state $w=w(a)$ \citep{1988NuPhB.302..645W,1988PhRvD..37.3406R}. In general, we can express equation (\ref{friedmann_density}) as

\begin{equation}
H^2=\frac{\kappa}{3}\left[\rho_{0m}\left(\frac{a_{0}}{a}\right)^3+\rho_{0r}\left(\frac{a_{0}}{a}\right)^4+\rho_{\phi}\right] \ ,
\end{equation}

\noindent where $\phi$ denotes the quintessence component. The evolution of $\rho_{\phi}$ can be obtained from the continuity equation

\begin{equation}
\label{continuity}
{\dot{\rho}}_{\phi}+3H(\rho_{\phi}+p_{\phi})=0 \ ,
\end{equation}

\noindent so that formally

\begin{equation}
\rho_{\phi}=\rho_{0\phi}\exp\left[-3\int_{a_{0}}^{a}\frac{1+w_{\phi}(a')}{a'}da'\right] \ ,
\end{equation}

\noindent where $w_{\phi} \equiv p_{\phi}/\rho_{\phi}$. The conserved energy density and pressure of the quintessence scalar field that appear in equation (\ref{continuity}) are defined as

\begin{equation}
\label{rho_OQ}
\rho_{\phi}=\frac{1}{2}{\dot{\phi}^2}+V(\phi) \ ,  
\end{equation}

\begin{equation}
\label{p_OQ}
p_{\phi}=\frac{1}{2}{\dot{\phi}^2}-V(\phi) \ , 
\end{equation}

\noindent where $V(\phi)$ is the potential in which the scalar field $\phi$ rolls.

Similarly, the evolution of the quintessence scalar field $\phi$ is given by the Klein-Gordon equation, obtained from equation (\ref{continuity}) substituting equations (\ref{rho_OQ}) and (\ref{p_OQ}),

\begin{equation}
\ddot{\phi}+3H\dot{\phi}+\frac{\partial V(\phi)}{\partial\phi}=0 \ .
\end{equation}

\noindent We note from equations (\ref{rho_OQ}) and (\ref{p_OQ}) that when the kinetic term ${\dot{\phi}^2/2}$ is negligible with respect to the potential term $V(\phi)$, then $w_{\phi} \rightarrow -1$ and the $\Lambda$CDM case is recovered.

In this paper, as potentials for minimally coupled quintessence models, we consider an inverse power law potential

\begin{equation}
\label{rp_potential}
V(\phi)=\frac{M^{4+\alpha}}{\phi^{\alpha}} \ ,
\end{equation}

\noindent the so called RP potential \citep{1988PhRvD..37.3406R}, as well as its generalization suggested by supergravity arguments \citep{1999PhLB..468...40B}, known as SUGRA potential, given by

\begin{equation}
\label{sugra_potential}
V(\phi)=\frac{M^{4+\alpha}}{\phi^{\alpha}}\exp(4\pi G \phi^2) \ ,
\end{equation}

\noindent where in both cases $M$ and $\alpha \ge 0$ are free parameters.

\subsection{Scalar-tensor theories} \label{EQ}

It is well possible that the quintessence scalar field might interact with other species. Here we consider the case in which $\phi$ interacts non-minimally with gravity \citep{1988NuPhB.302..645W,2000PhRvL..85.2236B} and we refer in particular to the extended quintessence (EQ) models described in \cite{2000PhRvD..61b3507P}, \cite{2005JCAP...12..003P} and \cite{2008PhRvD..77j3003P}. Scalar-tensor theories of gravity are generally described by the action

\begin{eqnarray}
\label{EQ_action}
S=\int d^4 x \sqrt{-g} \bigg[ \frac{1}{2\kappa} f(\phi,R) - \frac{\omega(\phi)}{2} \partial^{\mu} \phi \partial_{\mu} \phi - V(\phi) + \nonumber \\  
+ \ \mathcal{L}_{\rm fluid} \bigg] \ ,
\end{eqnarray}

\noindent where $R$ is the Ricci scalar, the function $f(\phi,R)$ specifies the coupling between the quintessence scalar field and the Ricci scalar, $\omega(\phi)$ and $V(\phi)$ specify the kinetic and potential terms respectively and the Lagrangian $\mathcal{L}_{\rm fluid}$ includes all the components but $\phi$. Here we assume for the sake of simplicity a standard form for the kinetic part, $\omega(\phi)=1$, and we define the coupling function as $f(\phi,R)=\kappa F(\phi)R$, where $F(\phi)$ is chosen to be 

\begin{equation}
\label{non-minimal coupling}
F(\phi)=\frac{1}{\kappa}+\xi (\phi^2 - {\phi_{0}^2}) \ .
\end{equation}

\noindent Here $\kappa \equiv 8 \pi G_{\ast}$, where $G_{\ast}$ represents the ``bare'' gravitational constant \citep{2001PhRvD..63f3504E}, which is in general different from the Newtonian constant $G$ and is set in such a way that locally $1/\kappa + \xi (\phi^2 - \phi_{0}^2) = 1 / 8 \pi G$ in order to match local constraints on General Relativity. The parameter $\xi$ represents the ``strength'' of the coupling. In particular we consider here a model with positive coupling $\xi > 0$ (EQp) and one with negative $\xi < 0$ (EQn). Note that a dependence on the sign is expected in this case \citep{2008PhRvD..77j3003P}. The limit of General Relativity is recovered when $\omega_{\rm JBD} \gg
1$, where

\begin{equation}
\omega_{\rm JBD}\equiv \frac{F(\phi)}{[\partial F(\phi)/\partial \phi]^2} \ .
\label{JBD}
\end{equation}

\noindent Stringent constraints for this quantity come from the Cassini mission
\citep{2003Natur.425..374B} on Solar System scales, where
$\omega_{\rm JBD0} > 4 \times 10^4$. However, it has been noted that
such constraints may not apply at cosmological scales
\citep{2005PhRvD..71l3526C} where complementary bounds, obtained
combining WMAP1 and 2dF large scale structure data, provide the less
tight limit of $\omega_{\rm JBD0} > 120$ at  $95$\% confidence level
\citep{2005PhRvD..71j4025A}.

In EQ models, we can define a conserved density and pressure for the scalar field, given by \citep{2002PhRvD..65l3505P}:

\begin{eqnarray}
\label{rho_EQ}
\rho_{\phi}=\frac{1}{2}{\dot{\phi}^2}+V(\phi)-3H\dot{F}(\phi) +3H^2\left[ \frac{1}{\kappa}-F(\phi)\right]  \ , 
\end{eqnarray}

\begin{eqnarray}  
\label{p_EQ}
p_{\phi}=\frac{1}{2}{\dot{\phi}^2}-V(\phi)+\ddot{F}(\phi)+2H\dot{F}(\phi)+
\nonumber \\ -(2\dot{H}+3H^2)\left[ \frac{1}{\kappa}-F(\phi)\right]
\ ,
\end{eqnarray}

\noindent respectively. The evolution of the scalar field follows from the Klein-Gordon equation

\begin{equation}
\label{klein-gordon_extended}
\ddot{\phi}+3H\dot{\phi}+\frac{\partial V(\phi)}{\partial\phi}=\frac{1}{2}\frac{\partial F(\phi)}{\partial \phi}R \ ,
\end{equation}

\noindent where the Ricci scalar is given by

\begin{equation}
\label{ricci}
R=6(\dot{H}+2H^2) \ .
\end{equation}

\noindent As in previous scenarios, equation (\ref{friedmann_density}) can be
again expressed as

\begin{equation}
H^2=\frac{\kappa}{3}\left[\rho_{0m}\left(\frac{a_{0}}{a}\right)^3+\rho_{0r}\left(\frac{a_{0}}{a}\right)^4+{\rho}_{\phi}\right] \ ,
\end{equation}
where $\rho_\phi$ is the conserved energy density defined in equation (\ref{rho_EQ}).

\noindent In this paper, as underlying potential for the extended quintessence models, we use the RP potential in equation (\ref{rp_potential}). Looking at equations (\ref{non-minimal coupling}), (\ref{rho_EQ}), (\ref{p_EQ}) we notice that minimally coupled quintessence is recovered for $\xi \rightarrow 0$. 

For an extensive linear treatment of EQ models we refer to
\citet{2008PhRvD..77j3003P}. Here we only recall for convenience that
EQ models behave like minimally coupled quintessence theories in
which, however, a time dependent effective gravitational interaction
is present. In particular, in the Newtonian limit, the Euler equation
for CDM can be written as

\begin{equation} 
\label{EQ_euler_cosmic} 
\nabla \dot{v}_m + H \nabla v_m + \frac{4 \pi \tilde{G} M_m
  \delta(0)}{a^2} = 0\ ,
\end{equation}

\noindent in terms of the cosmic time, where we have redefined the gravitational
parameter as

\begin{equation} 
\label{EQ_Gtilde_def} 
\tilde{G} = \frac{2 [ F + 2(\partial F/\partial \phi)^2]}{[ 2 F+3(\partial F/\partial \phi)^2]} \frac{1}{8 \pi F} \ .
\end{equation} 

\noindent The latter formalism is general for any choice of $F(\phi)$. For the
coupling here chosen and given by equation (\ref{non-minimal coupling}) we have 

\begin{equation}  
\label{tildeG}
\tilde{G} =  \frac{ \left[ \frac{ 1}{8 \pi G_\ast}  + (1 + 8 \xi) \xi
    \phi^2 - \xi {\phi_{0}^2 } \right]}{ \left[\frac{1}{8 \pi G_\ast}  +   (1  + 6 \xi)
    \xi \phi^2 - \xi {\phi_{0}^2 } \right]} \,  \frac{1}{\left[\frac{ 1}{G_\ast}  + 8 \pi
    \xi (\phi^2 - {\phi_{0}^2}) \right]}\ .
\end{equation}

\begin{figure}
\hbox{
 \epsfig{figure=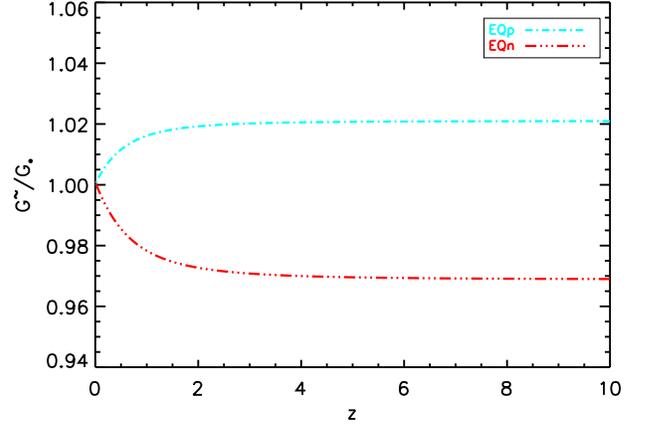,width=0.50\textwidth}
}
\caption{Correction to the gravity constant for the two extended quintessence models, EQp (cyan) and EQn (red), as expressed in equation (\ref{dG}). Note that the corrections are only within
the percent level.}
\label{dG_z}
\end{figure}

\noindent For small values of the coupling, that is to say $\xi \ll 1$, the latter expression becomes

\begin{equation}  
\frac{\tilde{G}}{\ G_\ast} \sim 1 - 8 \pi G_\ast \xi (\phi^2 - {\phi_{0}^2}) \ ,
\label{dG}
\end{equation}

\noindent which manifestly depends on the sign of the coupling $\xi$. We note that, since the derivative of the RP potential in equation (\ref{rp_potential}) with respect to $\phi$ is $\partial V(\phi)/ \partial \phi < 0$, we have $\phi^2 < {\phi_{0}^2}$. This leads to the behaviour of ${\tilde{G}}/{G_\ast}$ shown in Fig. \ref{dG_z}.

\subsection{Choice of the parameters} \label{choice}

As a reference model we use the $\Lambda$CDM model, adapted to the WMAP3 values \citep{2007ApJS..170..377S}, with the following cosmological parameters:

\begin{itemize}
\item matter density: $\Omega_{0m}=0.268$
\item dark energy density: $\Omega_{0\Lambda}=0.732$
\item baryon density: $\Omega_{0b}=0.044$
\item Hubble parameter: $h=0.704$
\item power spectrum normalization: $\sigma_{8}=0.776$
\item spectral index: $n_{s}=0.947$
\end{itemize}

\begin{figure}
\hbox{
 \epsfig{figure=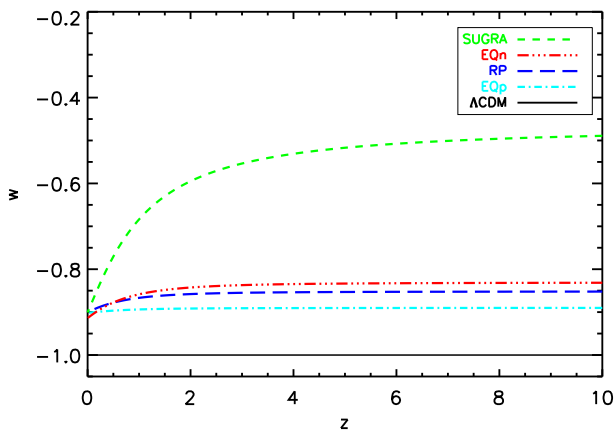,width=0.50\textwidth}
}
\caption{Redshift evolution of the equation of state parameter $w$ for the different cosmological models considered: $\Lambda$CDM (black), RP (blue), SUGRA (green), EQp (cyan), and EQn (red).}
\label{w_z}
\end{figure}

\begin{figure}
\hbox{
 \epsfig{figure=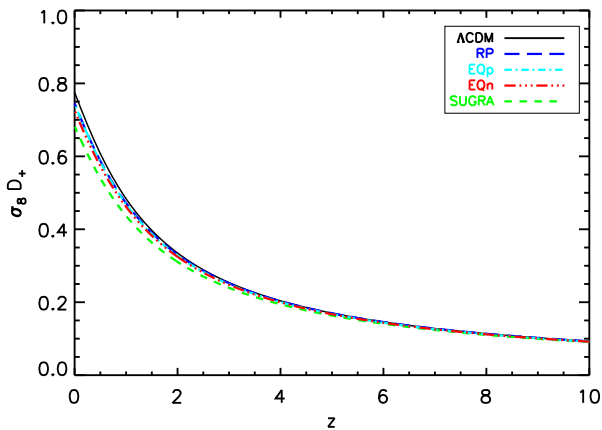,width=0.50\textwidth}
}
\caption{Redshift evolution of the growth factor $D_{+}$, normalized to the corresponding $\sigma_{8}$, for the different cosmological models considered: $\Lambda$CDM (black), RP (blue), SUGRA (green), EQp (cyan), and EQn (red).} 
\label{growth}
\end{figure}

We trimmed the parameters of the four dynamical dark energy models so that $w_0=w(0)\approx-0.9$ is the highest value still consistent with observational constraints in order to amplify the effects of dark energy. Fig. \ref{w_z} shows the evolution with redshift of $w$ in each cosmology. The parameters $\Omega_{0m}$, $\Omega_{0\Lambda}$, $\Omega_{0b}$, $h$, and $n_{s}$ are the same for all the models, but since we normalize all the dark energy models to CMB data from WMAP3, this leads to different values of $\sigma_{8}$ for the different cosmologies: 

\begin{equation}
\sigma_{8,\mathrm{DE}} = \sigma_{8,\mathrm{\Lambda CDM}} \frac{D_{+,\mathrm{\Lambda CDM}}(z_\mathrm{CMB})}{D_{+,\mathrm{DE}}(z_\mathrm{CMB})} \ ,
\label{sigmaDE}
\end{equation}

\noindent assuming $z_\mathrm{CMB}=1089$. This fact, along with the different evolution of the growth factor $D_{+}$ (shown in Fig. \ref{growth}), has an impact on structure formation. Table \ref{tab} lists the parameters chosen for the different cosmological models.

\begin{table}
\caption{Parameters chosen for the different cosmological models: $\alpha$ is the exponent of the inverse power law potential; $\xi$ is the coupling in the extended quintessence models; $w_{\rm JBD0}$ is the present value of the parameter introduced in equation (\ref{JBD}); $w_0$ is the present value of the equation of state parameter for dark energy; $\sigma_{8}$ is the normalization of the power spectrum as in equation (\ref{sigmaDE}).}
\begin{tabular}{|l|c|c|c|c|c|}
\hline
Model & $\alpha$ & $\xi$ & $w_{\rm JBD0}$ & $w_0$ & $\sigma_{8}$ \\ 
\hline
$\Lambda$CDM & --- & --- & --- & $-1.0$ & $0.776$ \\
RP & $0.347$ & ---  & --- & $-0.9$ & $0.746$ \\
SUGRA & $2.259$ & --- & --- & $-0.9$ & $0.686$ \\
EQp & $0.229$ & $+0.085$ & $120$ & $-0.9$ & $0.748$ \\ 
EQn & $0.435$ & $-0.072$ & $120$ & $-0.9$ & $0.729$ \\
\hline
\end{tabular}
\label{tab}
\end{table}

\section{Numerical simulations} \label{simulations}

In order to study the formation and evolution of large scale structures in
these different cosmological scenarios we use $N$-body + hydrodynamical
simulations done with the {\small{GADGET-3}} code
\citep{2001ApJ...549..681S,2005MNRAS.364.1105S}, 
which makes use of the entropy-conserving formulation of SPH
\citep{2002MNRAS.333..649S}. We extended the dark energy
implementation as described in \citet{2004A&A...416..853D} to allow the code to use
an external, tabulated Hubble function as well as a tabulated
correction to the gravity constant needed for the extended quintessence models.
The hydrodynamical simulations include radiative cooling,
heating by a uniform redshift-dependent UV background
\citep{1996ApJ...461...20H}, and a treatment of star formation and
feedback processes. The prescription of star formation we use is based on a
sub-resolution model to account for the multi-phase structure of the
interstellar medium (ISM), where the cold phase of the ISM is the
reservoir of star formation \citep{2003MNRAS.339..289S}. Supernovae (SNe) heat the
hot phase of the ISM and provide energy for evaporating some of the
cold clouds, thereby leading to self-regulation of the star formation
and an effective equation of state to describe its dynamics.
As a phenomenological extension of this feedback scheme,
\citet{2003MNRAS.339..289S} also included a simple model for galactic winds,
whose velocity, $v_w$, scales with the fraction $\eta$ of the Type II SN
feedback energy that contributes to the winds. The total
energy provided by Type II SN is computed by assuming that they are
due to exploding massive stars with mass $> 8 {\rm{M_{\odot}}}$ from a
\cite{1955ApJ...121..161S} initial mass function, with each SN
releasing $10^{51}$ ergs of energy. We have assumed $\eta=$ 0.5, yielding 
$v_w\simeq 340 {\rm{km s}}^{-1}$.

We simulated a cosmological box of size $(300 \ {\rm{Mpc}} \ h^{-1})^{3}$, resolved
with $(768)^{3}$ dark matter particles with a mass of 
$m_{DM} \approx 3.7 \times 10^{9} {\rm{M_{\odot}}} \ h^{-1}$ and the same amount of gas particles, having a mass of $m_{gas} \approx 7.3 \times 10^{8} {\rm{M_{\odot}}} \ h^{-1}$. As in \citet{2004A&A...416..853D}, we modified the initial conditions for the different dark energy scenarios 
adapting the initial redshift for the initial conditions in the dark energy
scenarios determined by the ratio of the linear growth factors $D_+(z)$,

\begin{equation}
  \frac{D_+(z_\mathrm{ini})}{D_+(0)}=
  \frac{D_\mathrm{+,\Lambda CDM}(z^\mathrm{ini}_\mathrm{\Lambda CDM})}
       {D_\mathrm{+,\Lambda CDM}(0)}\;.
\end{equation}
  
Additionally, the peculiar velocities of the particles are corrected according
to the new redshift to reflect a consistent application of the 
Zel'dovich approximation \citep{1970A&A.....5...84Z},

\begin{equation}
   \dot{x}(t)=\dot{D}_{+}(t)H(t)\nabla_q\Phi(\vec q)\;.
\end{equation}

\noindent Note that, unlike in previous works, here we do not use the approximation 
$\Omega_{m}^{0.6}$ for $\dot{D}_{+}(t)$ as this would lead to small inaccuracies in 
some of the dark energy scenarios. Finally we also correct the velocities
of the particles due to the changed displacement field at the new redshift
according to

\begin{equation}
  \vec v^\mathrm{\ ini}=\vec v_\mathrm{\Lambda CDM}^\mathrm{\ ini}\,
  \frac{\dot{D}_{+}(z_\mathrm{ini})\,H(z_\mathrm{ini})}
       {\dot{D}_\mathrm{+,\Lambda CDM}
        (z_\mathrm{\Lambda CDM}^\mathrm{ini})\,
        H_\mathrm{\Lambda CDM}
        (z_\mathrm{\Lambda CDM}^\mathrm{ini})} \ .
\end{equation}

\noindent Therefore, all simulations start from the same random phases, but the amplitude of the initial fluctuations is rescaled to satisfy the constraints given by CMB.

\begin{figure*}
\begin{center}
\includegraphics[width=0.4\textwidth]{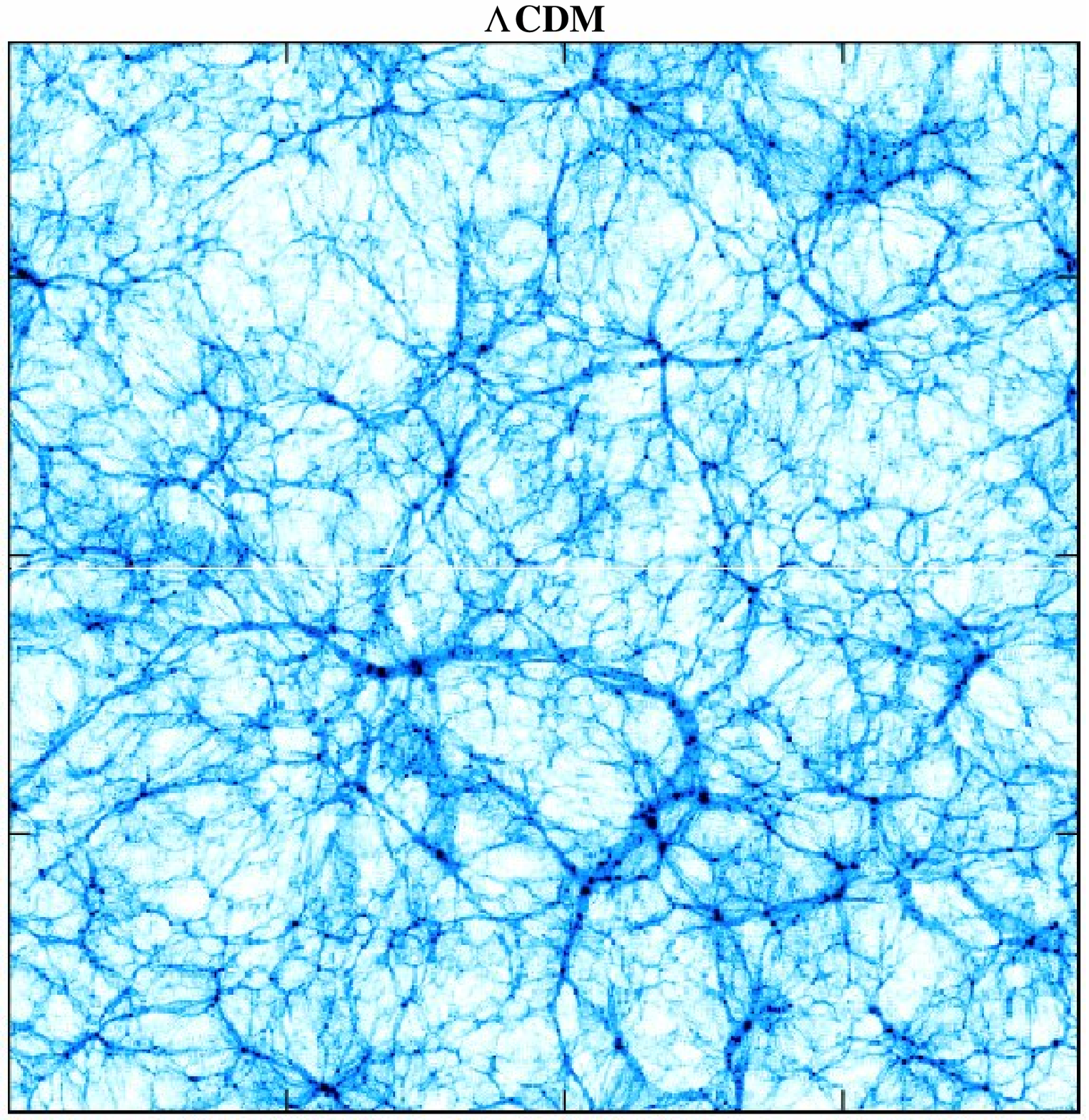}\\ 
\includegraphics[width=0.4\textwidth]{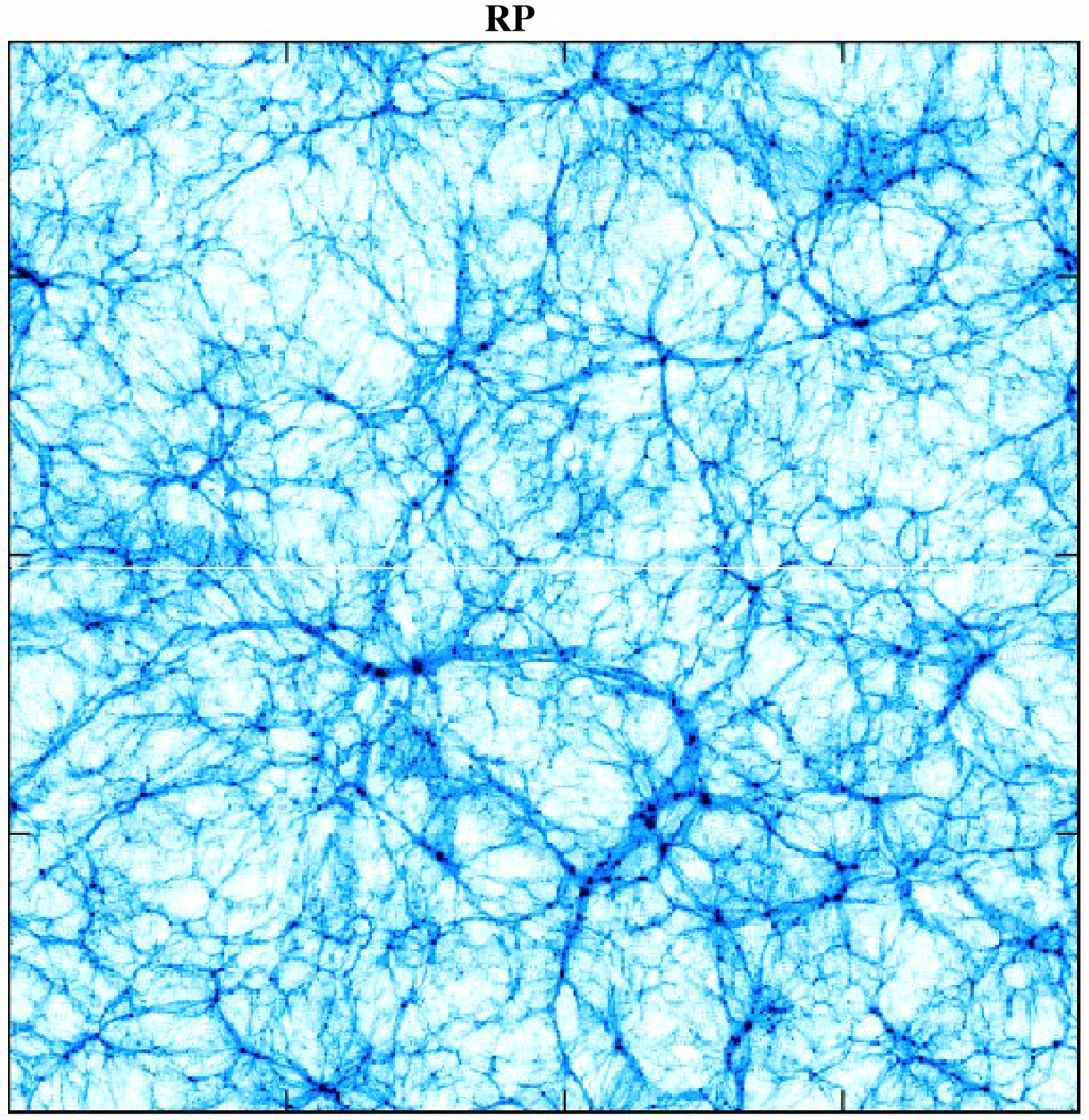}
\includegraphics[width=0.4\textwidth]{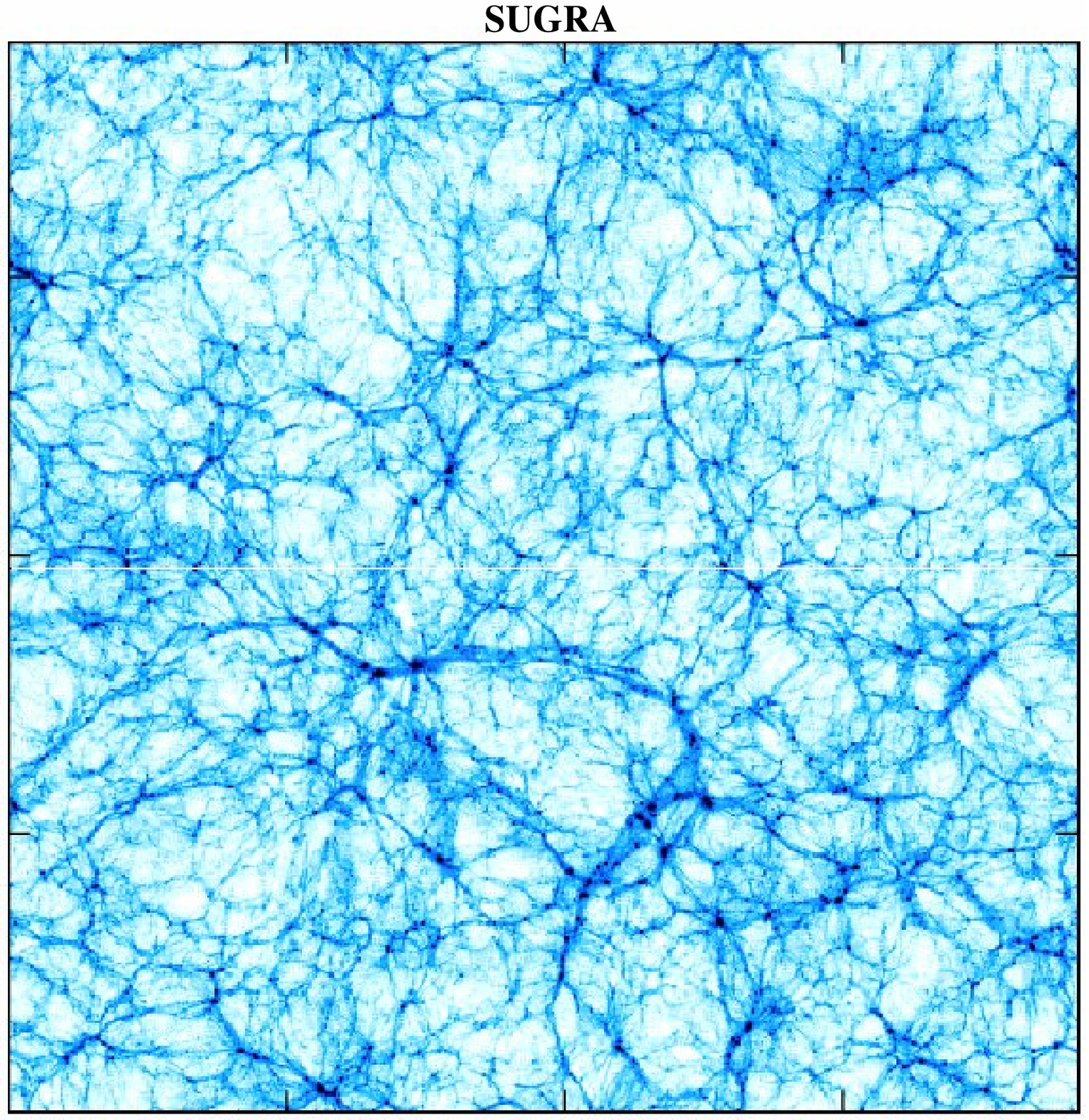}\\ 
\includegraphics[width=0.4\textwidth]{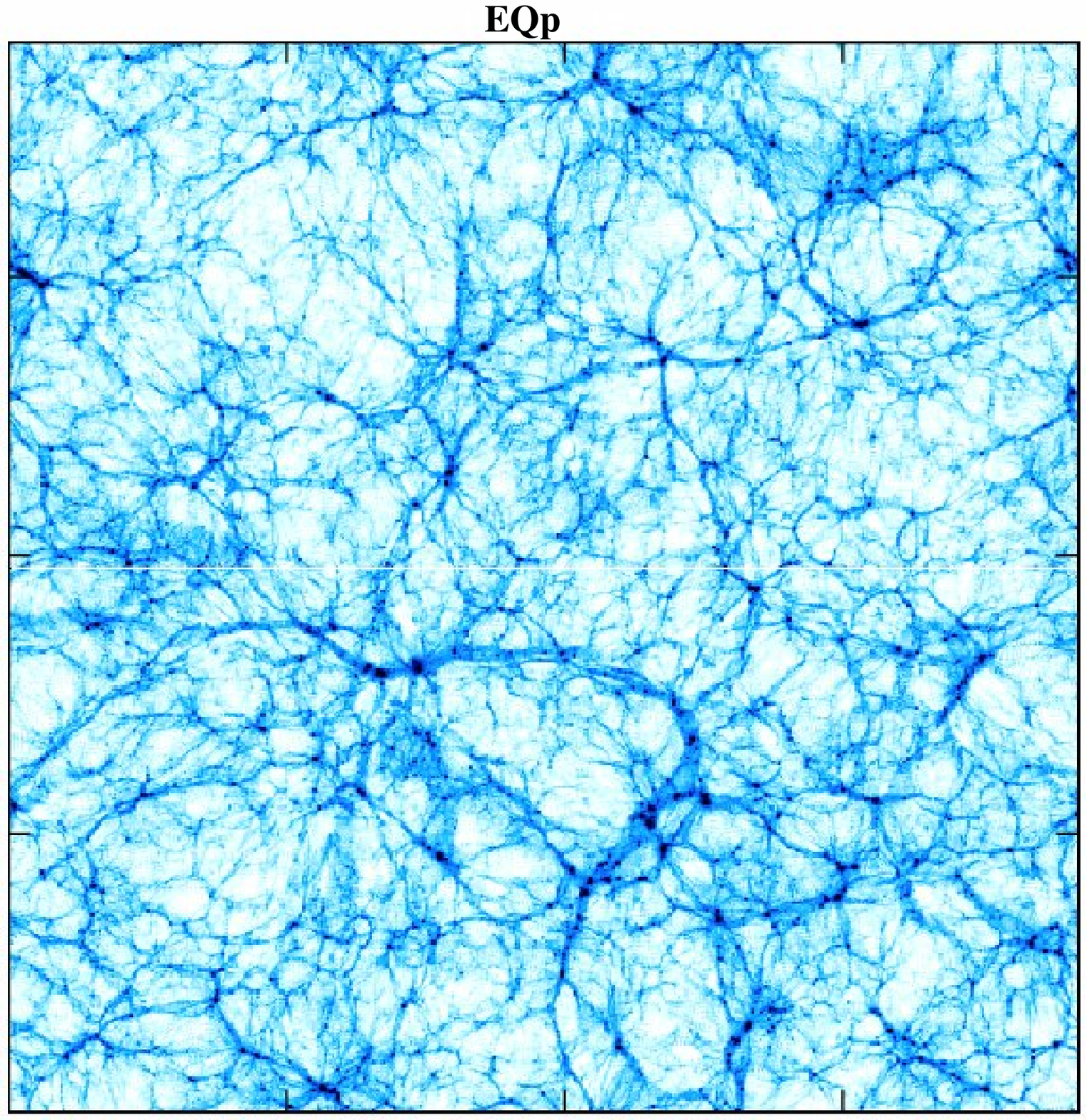}
\includegraphics[width=0.4\textwidth]{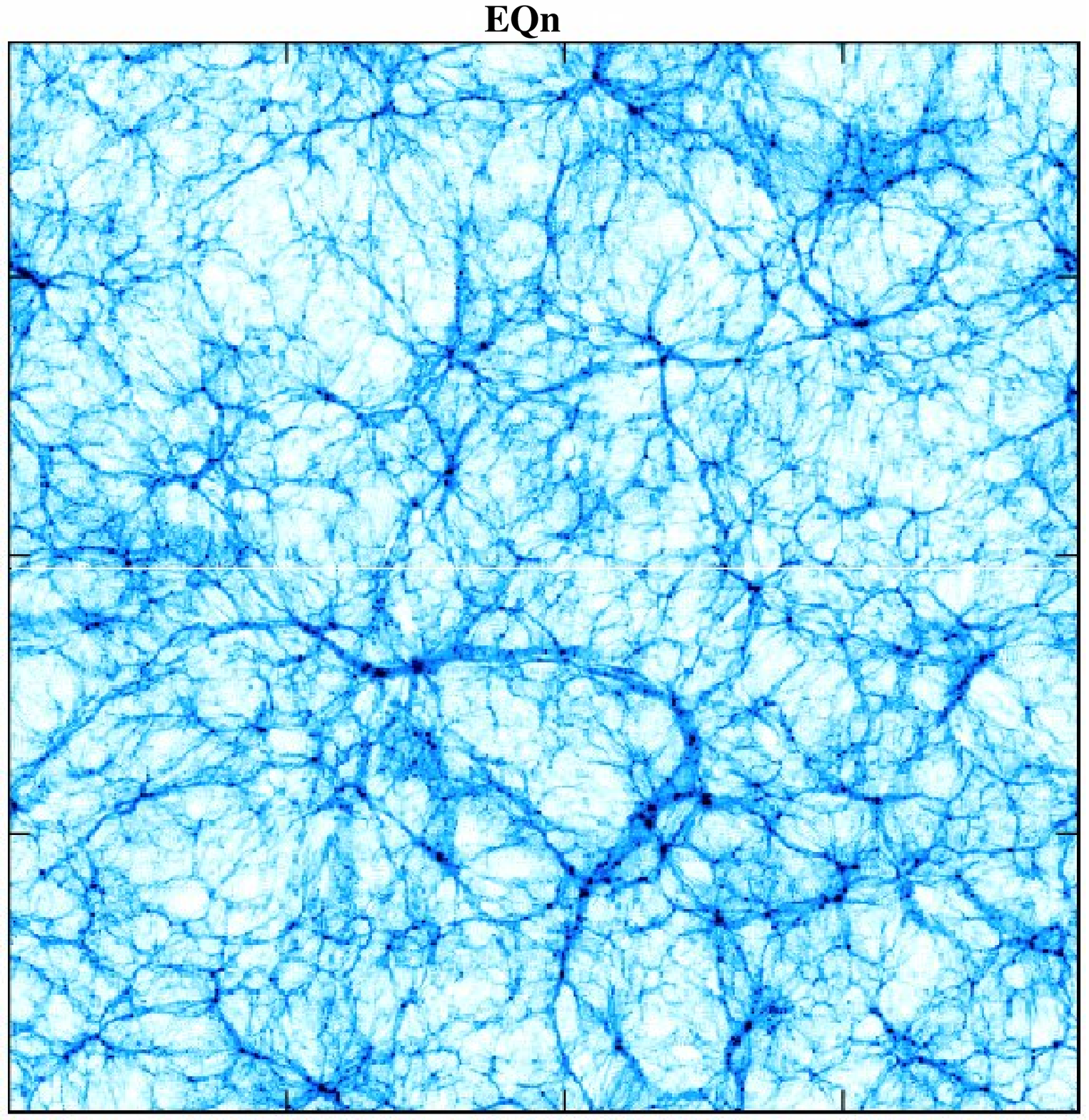}
\end{center}
\caption{Density slice of depth equal to $1/64$ of the box size 
through the whole simulation box for the five different models at $z=0$.}
\label{Maps}
\end{figure*}

In Fig. \ref{Maps} we show a density slice of depth equal to $1/64$ of the box size through the whole box for each of the five models considered at $z=0$. At first sight, we can see that the structures form in the same place in the different cosmologies since the initial phases are the same. Moreover, the differences among the models are small and cannot be seen with the eye; indeed, an accurate statistical analysis is needed to understand the properties of the objects in the different models.

\begin{figure}
\hbox{
 \epsfig{figure=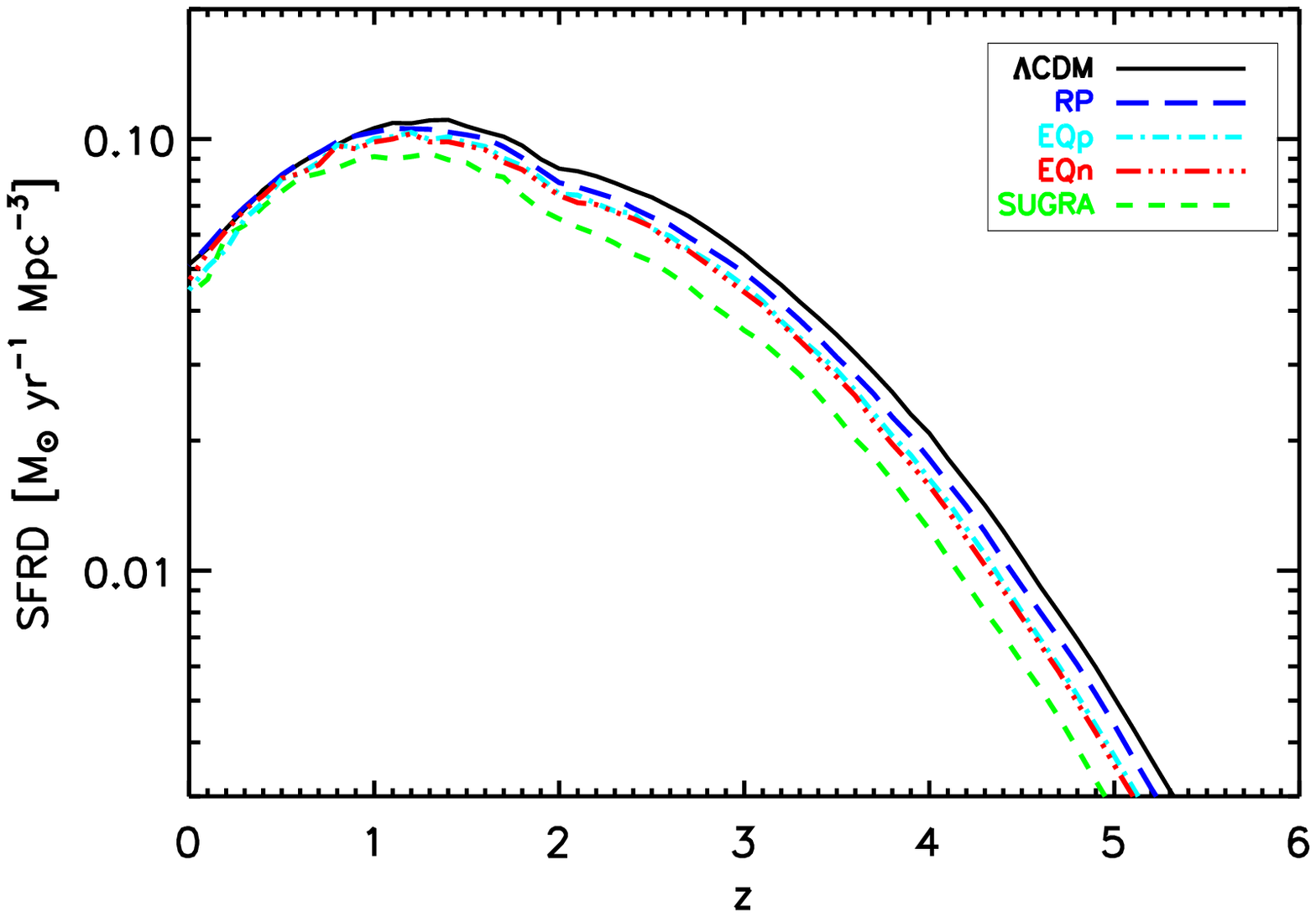,width=0.50\textwidth}
}
\caption{SFRD as a function of redshift for the $\Lambda$CDM model (black), RP (blue), SUGRA (green), EQp (cyan), and EQn (red) cosmologies.}
\label{SFRD}
\end{figure}

In Fig. \ref{SFRD} we plot the star formation rate density (SFRD) as a function of redshift for all the models considered. The SFRD in general follows the growth of the perturbations as shown in Fig. \ref{growth}.

Using the outputs of simulations, we extract galaxy clusters from the cosmological boxes, using the spherical overdensity criterion to define the collapsed structures. We take as halo centre the position of the most bound particle. Around this particle, we construct spherical shells of matter and stop when the total ({\it{i.e.}} dark matter plus gas plus stars) overdensity drops below $200$ times the {\it mean} (as opposed to {\it critical}) background density defined by $\Omega_{0m}\rho_{0c}$; the radius so defined is denoted with $R_{200m}$ and the mass enclosed in it as $M_{200m}$. We consider only halos that have $M_{200m}> 1.42 \times 10^{14}{\rm{M_{\odot}}}$. We select and study objects at three different redshifts, $z=0$, $z=0.5$, and $z=1$. For the following analysis, we also calculate for each cluster selected in this way the radius at which the overdensity drops below $200$ ($500$) times the {\it critical} background density and denote it with $R_{200}$ ($R_{500}$). The corresponding mass is indicated as $M_{200}$ ($M_{500}$). Just as a reference, the most massive object of all the simulations has $M_{200m} = 3.15 \times 10^{15}{\rm{M_{\odot}}}$. The number of clusters at each redshift is different for each cosmology: for example, the sample at $z=0$ is made up by 563 clusters in the $\Lambda$CDM cosmology, 484 in RP, 352 in SUGRA, 476 in EQp, and 431 in EQn. This fact directly reflects the different values of $\sigma_{8}$ and $D_{+}$ leading to differences in the formation history of the halos. No morphological selection has been made on the sample considered, so that clusters in very different dynamical state are included. Nevertheless, it is useful to define a quantitative criterion to decide whether a cluster can be considered relaxed or not because, in general, relaxed clusters have more spherical shapes, better defined centres and thus are more representative of the self-similar behaviour of the dark matter halos. We use a simple criterion similar to that introduced in \cite{2007MNRAS.381.1450N}: first of all we define $x_{off}$ as the distance between the centre of the halo (given by the most bound particle) and the barycentre of the region included in $R_{200m}$; then we define as relaxed the halos for which $x_{off}< 0.07 R_{200m}$.

\section{Mass function} \label{mass}

\begin{figure}
\hbox{
 \epsfig{figure=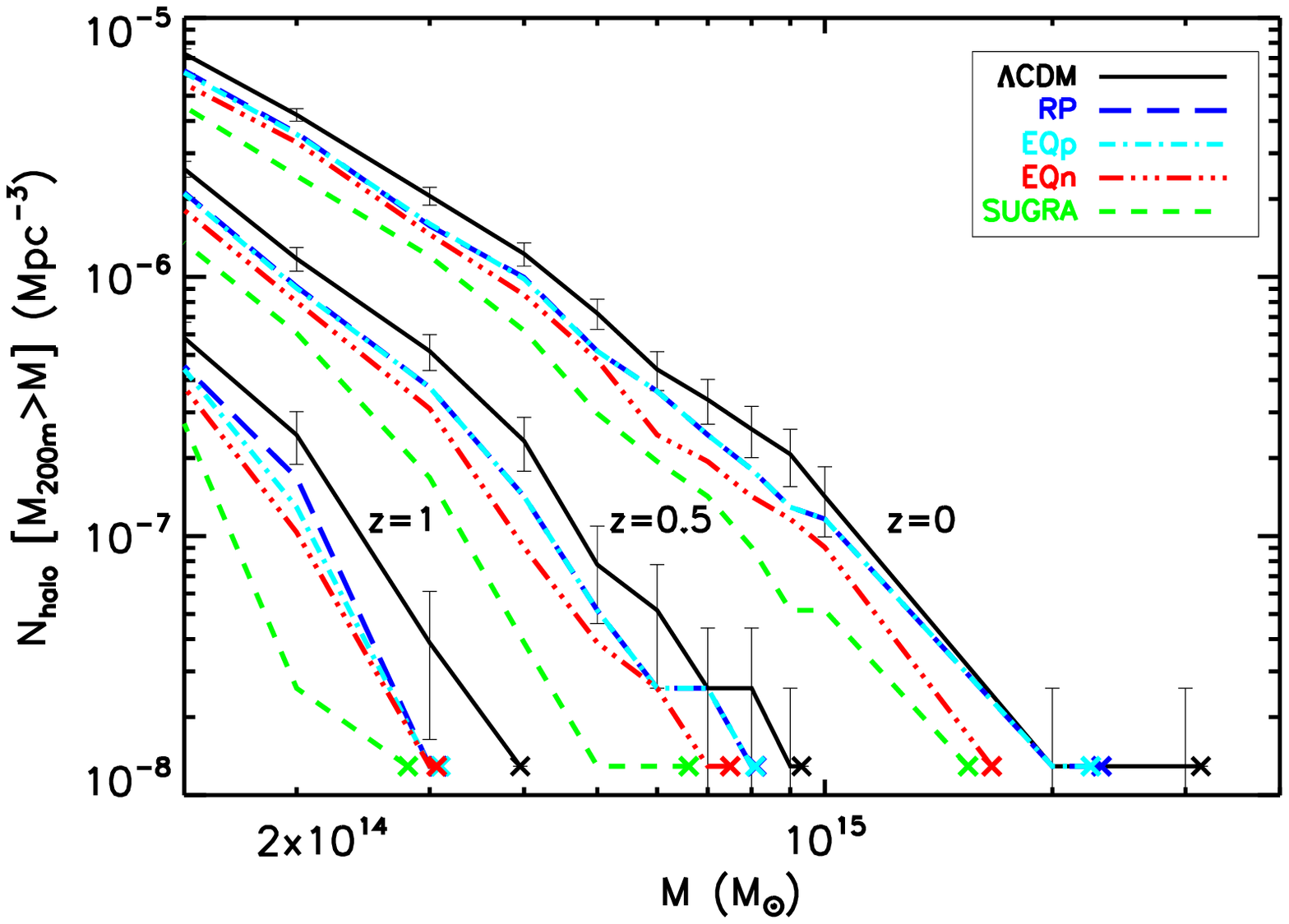,width=0.50\textwidth}
}
\hbox{
 \epsfig{figure=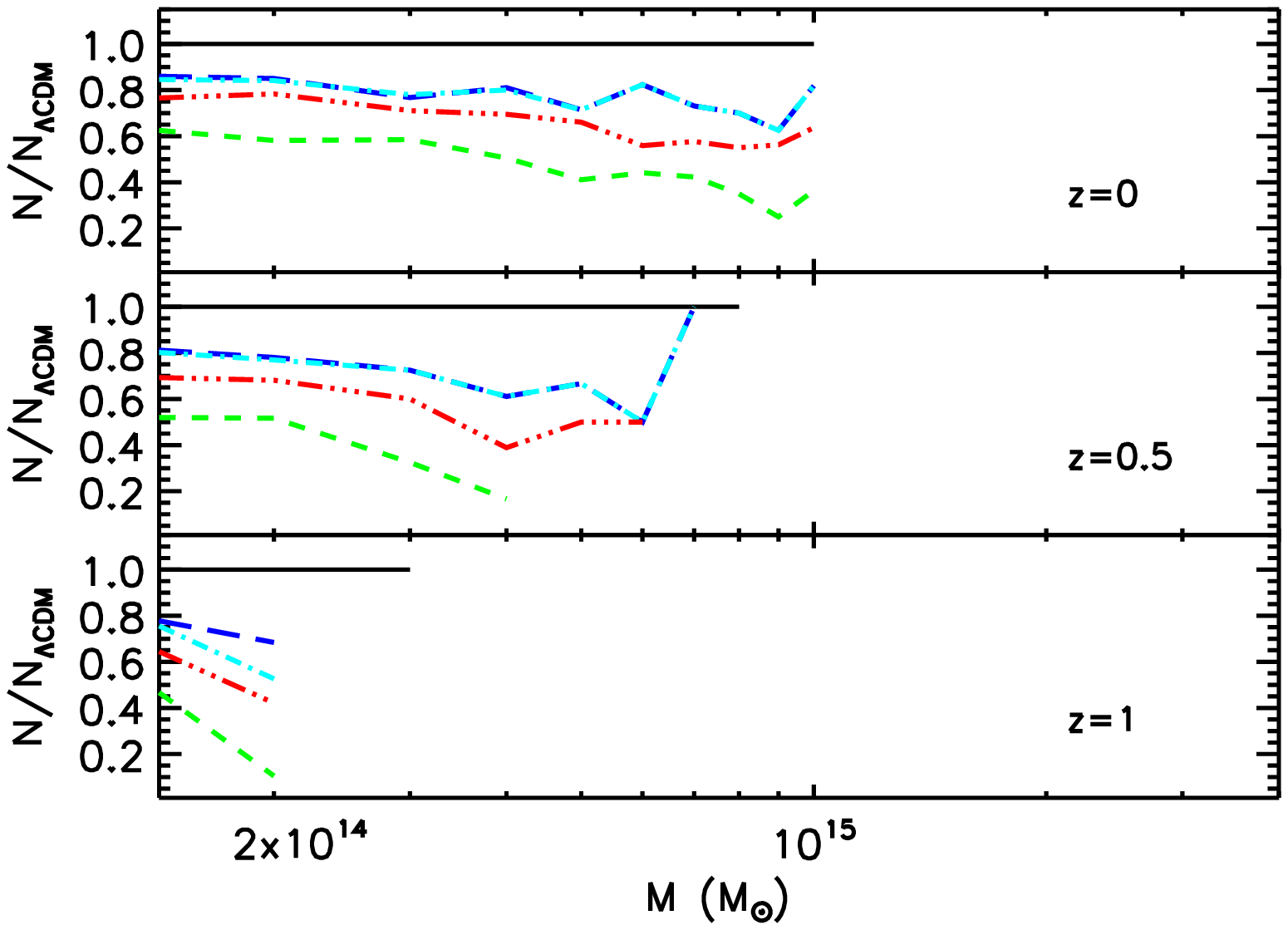,width=0.50\textwidth}
}
\caption{(Top panel) Cumulative mass function at $z=0$, $z=0.5$ and $z=1$ for the $\Lambda$CDM (black), RP (blue), SUGRA (green), EQp (cyan), and EQn (red) cosmologies. For each cosmological model the mass of the most massive object at each redshift is marked by a cross. Error bars (shown only for $\Lambda$CDM for clarity reasons) are Poissonian errors for the cluster number counts. (Bottom panel) Ratios between the mass functions for different dark energy cosmologies and the corresponding values for $\Lambda$CDM at $z=0$, $z=0.5$ and $z=1$.}
\label{mass_functions_alt}
\end{figure}

A standard way to use galaxy clusters as cosmological probe is the study of their mass function. Since the total mass of these objects is dominated by dark matter, it is a tracer of structure formation in different cosmological models.
In the top panel of Fig. \ref{mass_functions_alt} we plot the cumulative mass functions for the different cosmologies at three different redshifts: $z=0$, $z=0.5$ and $z=1$. This plot simply illustrates the number of halos per unit volume having a total mass greater than a given mass threshold. We can see that the shape and the properties of the mass functions are substantially the same at different redshifts (with the obvious exception of the maximum mass of the formed halos), with $\Lambda$CDM forming more clusters of a given mass with compared to the other cosmologies; SUGRA is the cosmology which forms fewer clusters, while RP, EQp and EQn lie in between, with RP and EQp being the closest to $\Lambda$CDM. This fact seems to directly reflect the redshift evolution of the equation of state parameter $w$ (see Fig. \ref{w_z}) and of the growth factor (see Fig. \ref{growth}), given the different value of $\sigma_{8}$ in the different models. Actually, for extended quintessence models, a positive value of the coupling $\xi$ leads to $\tilde{G} > {G_\ast}$ in the past, and vice versa for a negative $\xi$. Therefore, the linear density contrast is expected to be higher for EQp than for EQn. In a spherical collapse model like the Press-Schechter formalism \citep{1974ApJ...187..425P}, this implies a higher mass function for models with negative coupling ({\it{i.e.}} EQn) than for models with positive coupling ({\it{i.e.}} EQp), when all the other parameters are kept fixed. In our case, this effect is somehow mitigated by the different $\sigma_{8}$ used.

In the bottom panel of Fig. \ref{mass_functions_alt} we plot (always at $z=0$, $z=0.5$ and $z=1$) the ratios between the number of clusters in a given dark energy model with respect to the corresponding value in $\Lambda$CDM. For each cosmology, we consider only bins in which we have more than one object. The same results are summarized also in Table \ref{MF_tab}.
At $z=0$, RP, EQp, EQn, and SUGRA form $86$\%, $85$\%, $77$\%, and $63$\% the number of objects formed in $\Lambda$CDM, respectively. These numbers decrease with increasing redshift, reaching, at $z=1$, $78$\%, $76$\%, $64$\%, and $47$\% for RP, EQp, EQn, and SUGRA, respectively. This fact indicates that the differences in the formation history are more evident at high redshift. If we consider different mass bins at $z=0$, we see that the differences between $\Lambda$CDM and the other models are enhanced for very massive objects, in particular for SUGRA.

\begin{table}
\caption{Ratios between the number of clusters in the simulated volume for a given dark energy model with respect to ${\rm{N}}_{\Lambda {\rm{CDM}}}$ in the given $M_{200m}$ bin at different redshifts.}
\begin{tabular}{|l|c|c|r|r|r|r|}
\hline
$M_{200m}$ [$10^{14}{\rm{M_{\odot}}}$] & $z$ & ${\rm{N}}_{\Lambda {\rm{CDM}}}$ & RP \ & SUGRA & EQp & EQn \\ 
\hline
$>1.42$ & $0$ & $563$ & $0.86$ & $0.63$ \ \ \ & $0.85$ & $0.77$ \\
$>1.42$ & $0.5$ & $202$ & $0.81$ & $0.52$ \ \ \ & $0.80$ & $0.69$ \\
$>1.42$ & $1$ & $45$ & $0.78$ & $0.47$ \ \ \ & $0.76$ & $0.64$ \\
\\
$1.42-5$ & $0$ & $507$ & $0.88$ & $0.65$ \ \ \ & $0.86$ & $0.78$ \\
$5-10$ & $0$ & $45$ & $0.69$ & $0.42$ \ \ \ & $0.69$ & $0.67$ \\
$>10$ & $0$ & $11$ & $0.82$ & $0.36$ \ \ \ & $0.82$ & $0.64$ \\
\hline
\end{tabular}
\label{MF_tab}
\end{table}

Note that we have considered here minimally coupled models and scalar-tensor theories,
as illustrated in Sect. \ref{models}. Couplings with dark matter only, where, as in equation (\ref{EQ_euler_cosmic}), an
additional velocity-dependent term is present, have been shown
to lead to different results \citep{2011MNRAS.412L...1B}, increasing the
number of massive clusters at high redshift. Differences between these
sets of models have been illustrated in detail in \cite{2008PhRvD..77j3003P}.

In principle, if we can count all the clusters above a given mass threshold, or in a given mass bin, we can try to discriminate between different cosmologies just using cluster number counts coming from cosmological surveys. From a practical point of view, evaluating the mass of galaxy clusters requires the assumption of some hypotheses on their dynamical state, and in general it is not an easy task to perform. So it is better to consider cluster properties that are directly observable (like X-ray luminosity and temperature) in order to distinguish among different cosmologies. We discuss these topics in the next two sections.

\section{$L-T$ relation}  \label{LT}

Once we have analysed the general composition of our sample, we can now proceed with the study of the properties of the objects inside the sample. We recall that, when considering self-similar evolution of gravitational systems, we can derive simple scaling relations between their properties. The existence of such scaling relations is confirmed by observations, even if in general they have a different shape compared to the ones predicted by self-similarity, indicating an important role of some non-gravitational physics in the evolution of these systems. We use our hydrodynamical simulations in order to understand whether the baryon physics introduces any scale dependence that can break the self-similarity of the scaling relations. Since one of the aims of this work is to study whether there exist observable quantities that can be used to distinguish among the different cosmologies considered, we start studying the X-ray $L-T$ relation of our sample, also comparing it to observations to verify that the observed relation holds for our simulated objects too.

\begin{figure*}
\hbox{
 \epsfig{figure=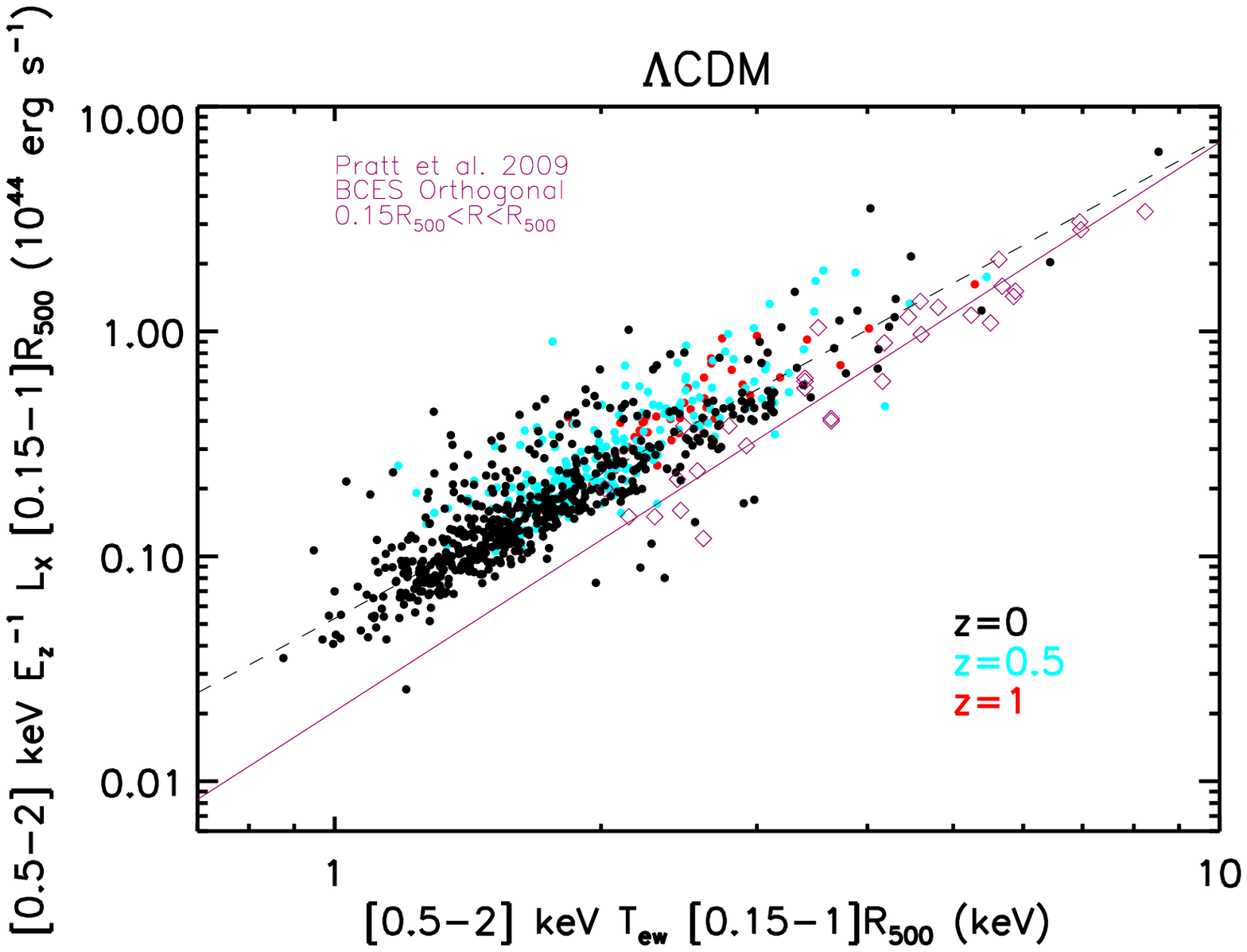,width=0.50\textwidth}
 \epsfig{figure=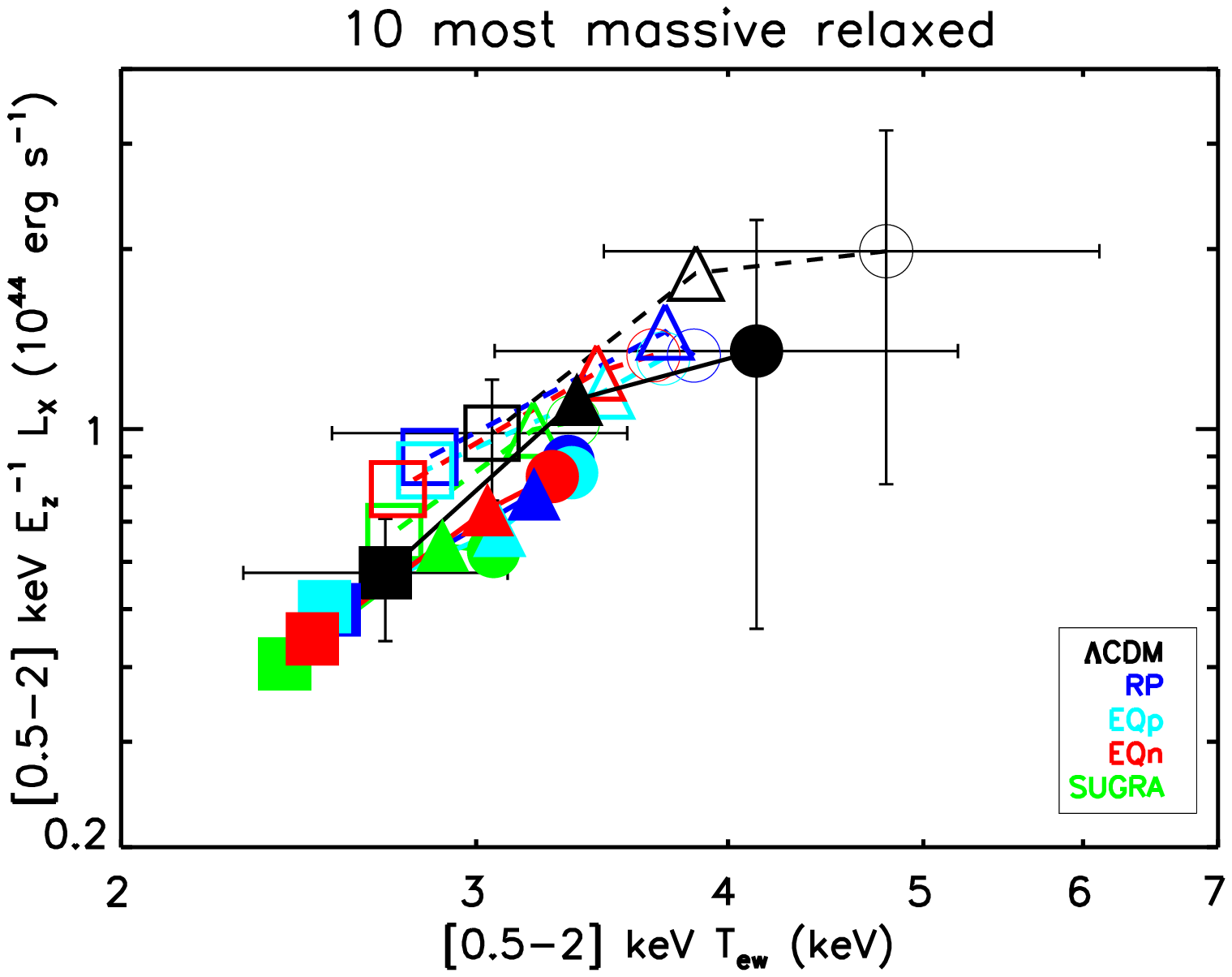,width=0.50\textwidth}
}
\caption{(Left panel) The X-ray $L-T$ relation in the [$0.5 - 2$] keV band, evaluated in the [$0.15 - 1$]$R_{500}$ region at $z=0$ (black), $z=0.5$ (cyan), and $z=1$ (red) for the $\Lambda$CDM cosmology. The dashed black line is the best-fit of our simulated data. The violet squares are a collection of observed data from Pratt et al. (2009), while the violet line is their best-fit of the same dataset. (Right panel) Redshift evolution of the mean luminosity and temperature in the [$0.5 - 2$] keV band for the ten most massive relaxed objects in the $\Lambda$CDM (black), RP (blue), SUGRA (green), EQp (cyan), and EQn (red) cosmologies. Circles refer to objects at $z=0$, triangles to objects at $z=0.5$ and squares to objects at $z=1$. Dashed lines and empty symbols indicate the evolution of the mean luminosity and temperature evaluated inside $R_{500}$, while solid lines and filled symbols refer to the same quantities evaluated in the [$0.15 - 1$]$R_{500}$ region. In both panels, the cosmological dependence is taken into account using the factor $E_{z}^{-1} \equiv H_{0} / H$ which multiplies the luminosity.}
\label{LT_evolution}
\end{figure*}

In order to do that, for each cluster we want to analyse we produce 2D maps of $(5 \ {\rm{Mpc}})^{2}$ size of the X-ray luminosity $L_{X}$ and emission-weighted temperature $T_{ew}$ in the [$0.5 - 2$] keV soft band. The latter is defined by

\begin{equation}
T_{ew} \equiv \frac{\int \Lambda(T) n^2 T dV}{\int \Lambda(T) n^2 dV} \ ,
\label{Tew}
\end{equation}

\noindent where $n$ is the gas density and $\Lambda(T)$ is the cooling function.

Then, for each object, we evaluate the total luminosity and the emission-weighted temperature in the region [$0.15 - 1$]$R_{500}$. We decide to cut the core for two reasons: first of all, despite the fact we use accurate physical models to describe the hydrodynamics of the simulations, still we do not include AGN feedback, so they are not optimized for the study of the central regions of the clusters; secondly, we have checked that cutting the core we obtain a lower dispersion of our data in the $L-T$ plane. We stress that despite excluding the central region of the clusters in our analysis we can still draw robust conclusions from a cosmological point of view, avoiding the effects of detailed physical processes which can affect the inner parts. Moreover this cut is often used in observations to avoid problems with cool-core emission that can lead to a deviation from the self-similar scaling relation. Having generated luminosity and temperature catalogues of our sample, we can proceed with the analysis of the $L-T$ relation. In the left panel of Fig. \ref{LT_evolution} we plot the $L-T$ relation at different redshifts ($z=0$, $z=0.5$ and $z=1$) for the $\Lambda$CDM cosmology. Here we correct the luminosity using $E_{z}^{-1} \equiv H_{0} / H$, which is a factor containing all the predicted dependence on the cosmology \citep[see {\it{e.g.}}][]{2004MNRAS.354..111E}. We can see that there are not substantial differences at the various redshifts, but in general at high redshift we lack clusters in the luminosity region below $10^{43} \rm{erg \ s}^{-1}$ and in the temperature region below $2$ keV. This fact can be explained as a selection effect in our sample: at high redshift, only more evolved (and thus more luminous and hotter) clusters are massive enough to be included in our sample. We also provide a fit to our points, fitting the linear relation between the logarithms of luminosity and temperature. We find a slope of $1.81$, which is slightly higher than the self-similar value of $1.5$ expected for the soft band considered. Finally we plot a collection of observed data at different redshifts compiled by \cite{2009A&A...498..361P}. The luminosities are taken exactly in the same way as we did, {\it{i.e.}} in the [$0.15 - 1$]$R_{500}$ region and in the [$0.5 - 2$] keV band, while they use spectroscopically determined temperatures \citep[see the details in][]{2009A&A...498..361P}. The slope of their best-fit relation is $2.53 \pm 0.16$, steeper than what we found. Despite the difference in the slope, we can see that in the high-temperature/high-luminosity region where we have a sufficient number of both observed and simulated objects, the agreement is very good. In any case, we stress that a direct comparison between simulations and observations is not the main target of this work. Here, we just want to show that our simulated clusters lie in a region in the $L-T$ plane which is the same as the observed objects. Regarding the differences we find in the low-temperature/low-luminosity region, we stress that it is not due to overcooling in the simulations, since we are cutting the core; more likely, this region is populated by objects with lower mass, for which the detailed physical processes acting in the inner regions ({\it{e.g.}} AGN feedback) have important effects also on the overall properties of the clusters \cite[see {\it{e.g.}}][]{2008ApJ...687L..53P}. 

In the right panel of Fig. \ref{LT_evolution} we plot the evolution with redshift of the mean luminosity and temperature in the different cosmologies.
We consider only the relaxed clusters at $z=0$, $z=0.5$ and $z=1$. Then, for each cosmology, we select the ten most massive objects at each redshift, using $M_{200}$ for this selection. Actually, at $z=1$ for the SUGRA model we only have six relaxed clusters, and we consider all of them. At this point, at each redshift, we evaluate the mean luminosity and temperature of the selected objects both in the region inside $R_{500}$ and in the region [$0.15 - 1$]$R_{500}$. We find that cutting the core results in both a lower mean luminosity and lower mean emission-weighted temperature. As a general trend, either including or cutting the core, both the mean luminosity and temperature increase with decreasing redshift, independently of the cosmological model. This is in somehow expected, since at late cosmic time the clusters are more evolved, and thus hotter and more luminous. The differences in the values of mean luminosity and temperature among the different cosmologies reflect the different histories experienced by objects in different dark energy environments, substantially following the mass function.

\section{X-ray observable functions}  \label{functions}

\begin{figure*}
\hbox{
 \epsfig{figure=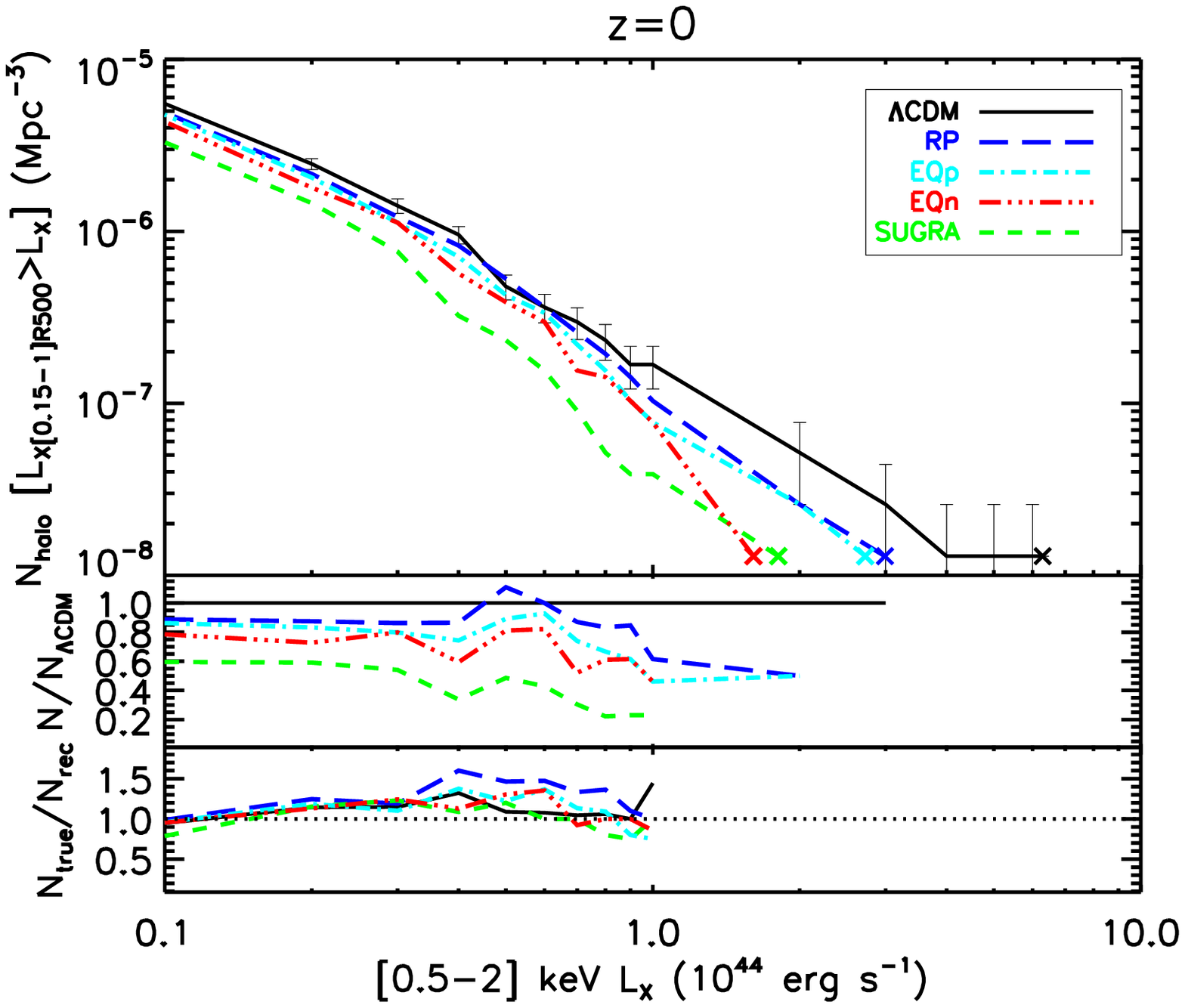,width=0.50\textwidth}
 \epsfig{figure=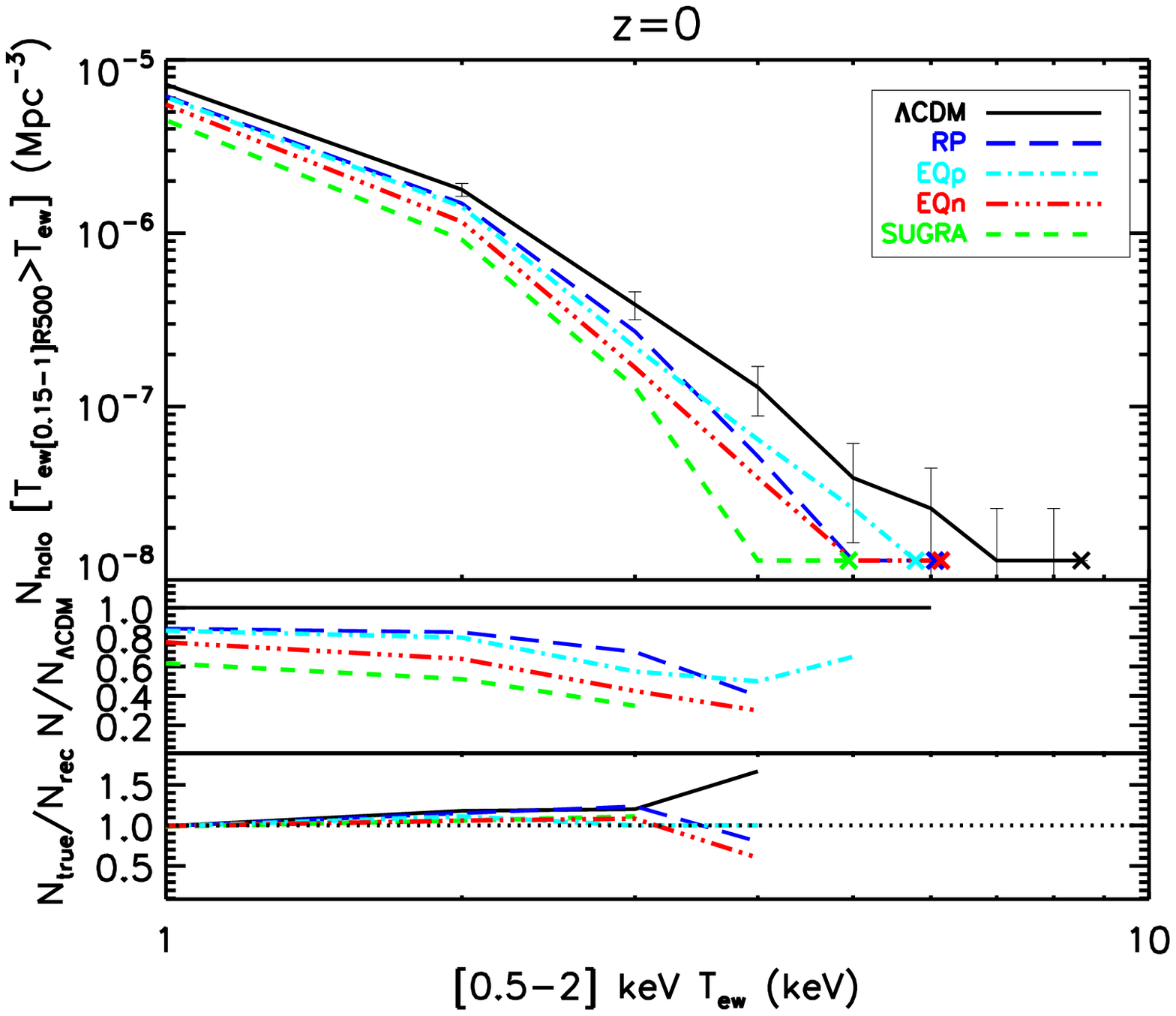,width=0.50\textwidth}
}
\hbox{
}
\hbox{
}
\caption{(Left panel) The X-ray luminosity function (XLF) in the [$0.5 - 2$] keV band, evaluated in the [$0.15 - 1$]$R_{500}$ region at $z=0$ for the $\Lambda$CDM (black), RP (blue), SUGRA (green), EQp (cyan), and EQn (red) cosmologies. For each cosmological model the luminosity of the object with the highest luminosity is marked by a cross. Error bars (shown only for $\Lambda$CDM for clarity reasons) are Poissonian errors for the cluster number counts. In the middle panel the ratios between the luminosity functions for different dark energy cosmologies and the corresponding values for $\Lambda$CDM are shown. In the bottom panel we plot the ratio between the luminosity functions shown in the top panel and the ones recovered by applying the $L-M$ relation at $z=0$ for the $\Lambda$CDM cosmology to the mass function of each dark energy model. (Right panel) The same as in the left panel, but for the X-ray temperature function (XTF).} 
\label{XLF_XTF}
\end{figure*}

Using the same maps built to study the X-ray $L-T$ relation, we can also analyse the X-ray luminosity function (XLF) and the X-ray temperature function (XTF) of our samples. Since the samples are mass selected (see Sect. \ref{mass}), only the mass functions we have shown before can be considered complete. XLFs and XTFs in a sense reflect the mass functions, but cannot be considered complete for the selection effect discussed in the previous section. This means that at higher redshift, we are missing more and more clusters in the low-luminosity region of the XLF and in the low-temperature region of the XTF. We show in Fig. \ref{XLF_XTF} the cumulative XLFs and XTFs of our sample at $z=0$. We cut the plots at $0.1 \times 10^{44} \rm{erg \ s}^{-1}$ and $1$ keV in order to be as complete as possible also in the low-luminosity and low-temperature regions. 
In the left panel of Fig. \ref{XLF_XTF} we show the cumulative luminosity function. In the middle panel of the same figure we plot the ratios between the number of clusters in a given dark energy model with respect to $\Lambda$CDM in every luminosity bin. As in the case of the mass function, for each cosmology, we consider only bins in which we have more than one object. 
The results for three luminosity bins are also summarized in Table \ref{Lx_Tew_tab}. In general, despite some noisy oscillations, the ratio is decreasing with increasing luminosity. Nevertheless, in the range between $0.5$ and $1 \times 10^{44} \rm{erg \ s}^{-1}$ it increases and in three models out of four the number of objects is equal or even larger than in $\Lambda$CDM. This effect seems to be statistically significant in particular for RP. In any case, by looking only at very luminous objects, the differences with $\Lambda$CDM are significant for all models.

In the right panel of Fig. \ref{XLF_XTF} we show the same as in the left panel, but for the cumulative temperature function (see also Table \ref{Lx_Tew_tab}). In this case, the decrease of the ratio with increasing temperature is evident in all the dynamical dark energy cosmologies. Going from objects in the range between $1$ and $3$ keV to objects with temperatures higher than $3$ keV, RP goes from $87$\% to $70$\%, SUGRA from $64$\% to $33$\%, EQp from $86$\% to $57$\%, and EQn from $78$\% to $43$\%.

\begin{table}
\caption{Ratios between the number of clusters in the simulated volume for a given dark energy model with respect to ${\rm{N}}_{\Lambda {\rm{CDM}}}$ in the given $L_{X}$, $T_{ew}$, $M_{gas500}$, and $Y_{X500}$ bin.}
\begin{tabular}{|l|c|r|r|r|r|r|}
\hline
 $z=0$ & ${\rm{N}}_{\Lambda {\rm{CDM}}}$ & RP \ & SUGRA & EQp & EQn \\
\hline
$L_{X}$ [$10^{44} \rm{erg \ s}^{-1}$] & & & & & \\ 
\hline
$0.1 - 0.5$ & $391$ & $0.87$ & $0.61$ \ \ \ & $0.86$ & $0.78$ \\
$0.5 -1$ & $24$ & $1.38$ & $0.63$ \ \ \ & $1.13$ & $1.00$ \\
$> 1$ & $13$ & $0.62$ & $0.23$ \ \ \ & $0.46$ & $0.46$ \\
\hline
$T_{ew}$ [keV] & & & & & \\ 
\hline
$1-3$ & $528$ & $0.87$ & $0.64$ \ \ \ & $0.86$ & $0.78$ \\
$>3$ & $30$ & $0.70$ & $0.33$ \ \ \ & $0.57$ & $0.43$ \\
\hline
$M_{gas500}$ [$10^{13}{\rm{M_{\odot}}}$] & & & & & \\ 
\hline
$1-5$ & $347$ & $0.85$ & $0.58$ \ \ \ & $0.81$ & $0.76$ \\
$> 5$ & $14$ & $0.50$ & $0.21$ \ \ \ & $0.43$ & $0.50$ \\
\hline
$Y_{X500}$ [$10^{13}{\rm{M_{\odot}} \ keV}$ ]& & & & & \\ 
\hline
$1-5$ & $392$ & $0.87$ & $0.63$ \ \ \ & $0.82$ & $0.79$ \\
$5-10$ & $55$ & $0.98$ & $0.49$ \ \ \ & $0.96$ & $0.96$ \\
$> 10$ & $29$ & $0.69$ & $0.38$ \ \ \ & $0.62$ & $0.66$ \\
\hline
\end{tabular}
\label{Lx_Tew_tab}
\end{table}

In general, we see that the relative trend among the different cosmologies shown by the mass functions at $z=0$ is almost preserved in the XLFs and XTFs: in a given mass, luminosity and temperature bin, $\Lambda$CDM forms more clusters than the other cosmologies do (except for RP in a luminosity bin, as noted before). On the other hand, SUGRA is the cosmological model that forms fewer clusters in each bin. EQp and EQn lie in between. This finding is confirmed by the bottom panels of Fig. \ref{XLF_XTF} where we show the ratios between the XLFs and XTFs plotted in the top panels and the ones recovered by applying the $L-M$ relation at $z=0$ for the $\Lambda$CDM cosmology to the mass functions of each dark energy model. This is done to disentangle the differences in the XLFs and XTFs due to a different mass function and the ones due to baryon physics.
 The fact that the subsample considered in the right panel of Fig. \ref{LT_evolution} reproduces the XLF and XTF of Fig. \ref{XLF_XTF} also seems to indicate that relaxed and massive objects are still a good representation of the whole sample. The general trend of the mass, luminosity and temperature functions seems to reflect the evolution with redshift of the dark energy equation of state parameter $w=w(z)$, as we showed in Fig. \ref{w_z}. $\Lambda$CDM tends to form massive clusters earlier than the other cosmologies, thus giving a larger number of evolved ({\it{i.e.}} with high luminosity and temperature) objects at $z=0$. The XTF seems to better reflect the mass function, while the XLF is more influenced by baryonic physics, as we can clearly see from the behaviour of the RP cosmology. So, in principle, we can try to distinguish among different cosmologies by building the XTF of a sample of galaxy clusters. The problem is that if we check, for example, the sample from \cite{2009A&A...498..361P}, there are very few clusters in the temperature range we have considered for our XTF. Being an X-ray selected sample, in general they have a higher temperature compared with our simulated objects, and so it is not easy to directly compare our results with their observational data. 

In order to check whether other proxies could better trace the formation history of structures, we also analysed the X-ray $M_{gas500}$ and $Y_{X500}$ functions. $M_{gas500}$ is defined simply by the mass of X-ray emitting gas contained in $R_{500}$, while $Y_{X500} = M_{gas500} \times T_{ew}$, where $T_{ew}$ is evaluated in the [$0.15 - 1$]$R_{500}$ region. We see from Table \ref{Lx_Tew_tab} that, for $M_{gas500} > 5 \times 10^{13}{\rm{M_{\odot}}}$, $M_{gas500}$ is in principle a very powerful tool to distinguish between different cosmologies. In fact, all the models form at most $50$\% the number of objects formed by $\Lambda$CDM, and, since $M_{gas500}$ is quite an easy quantity to estimate from observations, it should be possible to rule out some models just by studying the $M_{gas500}$ function. The quantity $Y_{X500}$ does not seem to be as good as $M_{gas500}$, since the differences between $\Lambda$CDM and the other models are less pronounced, and also the behaviour in the different bins is not so smooth. 

It is interesting to evaluate the volume that a cluster survey
must cover to be able to discriminate using the local ({\it{i.e.}} at $z=0$)
cluster counts among the different dark energy models here
considered. For that we assume Poissonian error bars and consider a
$3 \sigma$ level. Using the mass function with a threshold of $5 \times
\ 10^{14}\rm{M_{\odot}}$, we find that cosmological volumes larger than
$1.6 \times 10^7 ({\rm{Mpc}} \ h^{-1})^{3}$ are sufficient to distinguish between SUGRA and
$\Lambda$CDM, while $6.4 \times 10^7 ({\rm{Mpc}} \ h^{-1})^{3}$ are required for EQn and $9.1 \times 10^7 ({\rm{Mpc}} \ h^{-1})^{3}$ are required for RP and EQp. Considering
the XLF (with a threshold of $0.5 \times 10^{44} \rm{erg \ s}^{-1}$),
larger surveys are required: in fact volumes larger than $4.3 \times
10^7$, $3.4 \times 10^8$, $1 \times 10^9$, and $1.3 \times 10^{9} ({\rm{Mpc}} \ h^{-1})^{3}$ are necessary to discriminate among SUGRA, EQn, EQp, and RP and
$\Lambda$CDM, respectively. The situation is better when the XTF (with
a threshold of $3$ keV) is used: predictions for the $\Lambda$CDM model
are different at 3$\sigma$ level with respect to the ones for SUGRA,
EQn, EQp, and RP, when volumes as large as $2.7 \times 10^7$, $4.3
\times 10^7$, $6.4 \times 10^7$, and $1.7 \times 10^8 ({\rm{Mpc}} \ h^{-1})^{3}$ are
considered, respectively. This fact confirms the importance of XLF/XTF in tracing
the number counts in a given cosmology, and again that the XTF is a
better quantity to be used in that kind of studies, if compared to the
XLF. We recall that we are not considering any selection function on
XLF/XTF, being a proper treatment of any observational approach in
defining an XLF/XTF beyond the purpose of the present work. If we move to $z=1$, using the mass function with a threshold of $1.42 \times 10^{14}\rm{M_{\odot}}$, $2.7 \times
10^7 ({\rm{Mpc}} \ h^{-1})^{3}$ are still sufficient to distinguish between SUGRA and $\Lambda$CDM, while EQn, EQp, and RP need $6.4 \times 10^7 ({\rm{Mpc}} \ h^{-1})^{3}$, $1.7 \times 10^8 ({\rm{Mpc}} \ h^{-1})^{3}$, and $2.2 \times 10^8 ({\rm{Mpc}} \ h^{-1})^{3}$ to be distinguished from $\Lambda$CDM, respectively.
Larger cosmological boxes and larger observational samples with higher resolution and sensitivity ({\it{i.e.}} lower flux limit), such as, {\it{e.g.}}, the one expected with eROSITA \citep{2007SPIE.6686E..36P} and WFXT \citep{2009astro2010S..90G}, can provide better answers to the question.

\section{The baryon fraction}  \label{bias}

In this section we focus on the baryon fraction $f_{bar}=f_{star}+f_{gas}$ of our simulated galaxy clusters, where $f_{star} \equiv M_{star}/M_{tot}$ and $f_{gas} \equiv M_{gas}/M_{tot}$. Since we are considering galaxy clusters in a cosmological context, it is better to re-express the star fraction $f_{star}$, the gas fraction $f_{gas}$, and the total baryon fraction $f_{bar}$ in units of the cosmic baryon fraction $\Omega_{0b}/\Omega_{0m}=0.164$ adopted in these simulations. To do this we introduce the quantities 

\begin{equation}
b_{star} \equiv \frac{f_{star}}{\Omega_{0b}/\Omega_{0m}} \ ; 
\ \ b_{gas} \equiv \frac{f_{gas}}{\Omega_{0b}/\Omega_{0m}} \ ; 
\ \ b_{bar} \equiv \frac{f_{bar}}{\Omega_{0b}/\Omega_{0m}} \ ,
\label{depletion}
\end{equation}

\noindent and indicate them as star, gas and baryon depletion parameter, respectively.
In this section we analyse the dependence of these quantities on mass, redshift and distance from the centre of the object considered, as well as on the underlying cosmology.

\begin{figure}
\hbox{
 \epsfig{figure=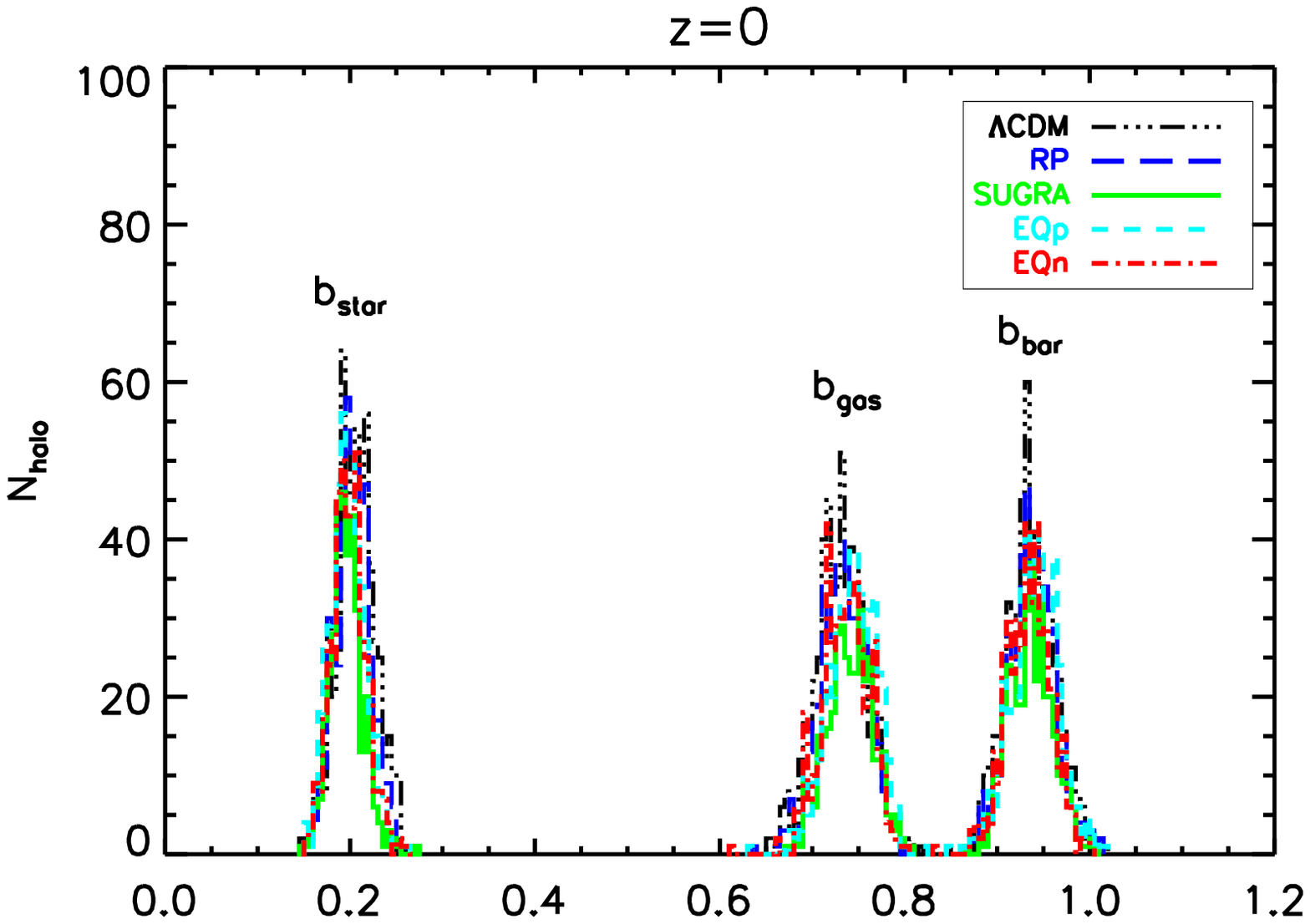,width=0.50\textwidth}
}
\hbox{
 \epsfig{figure=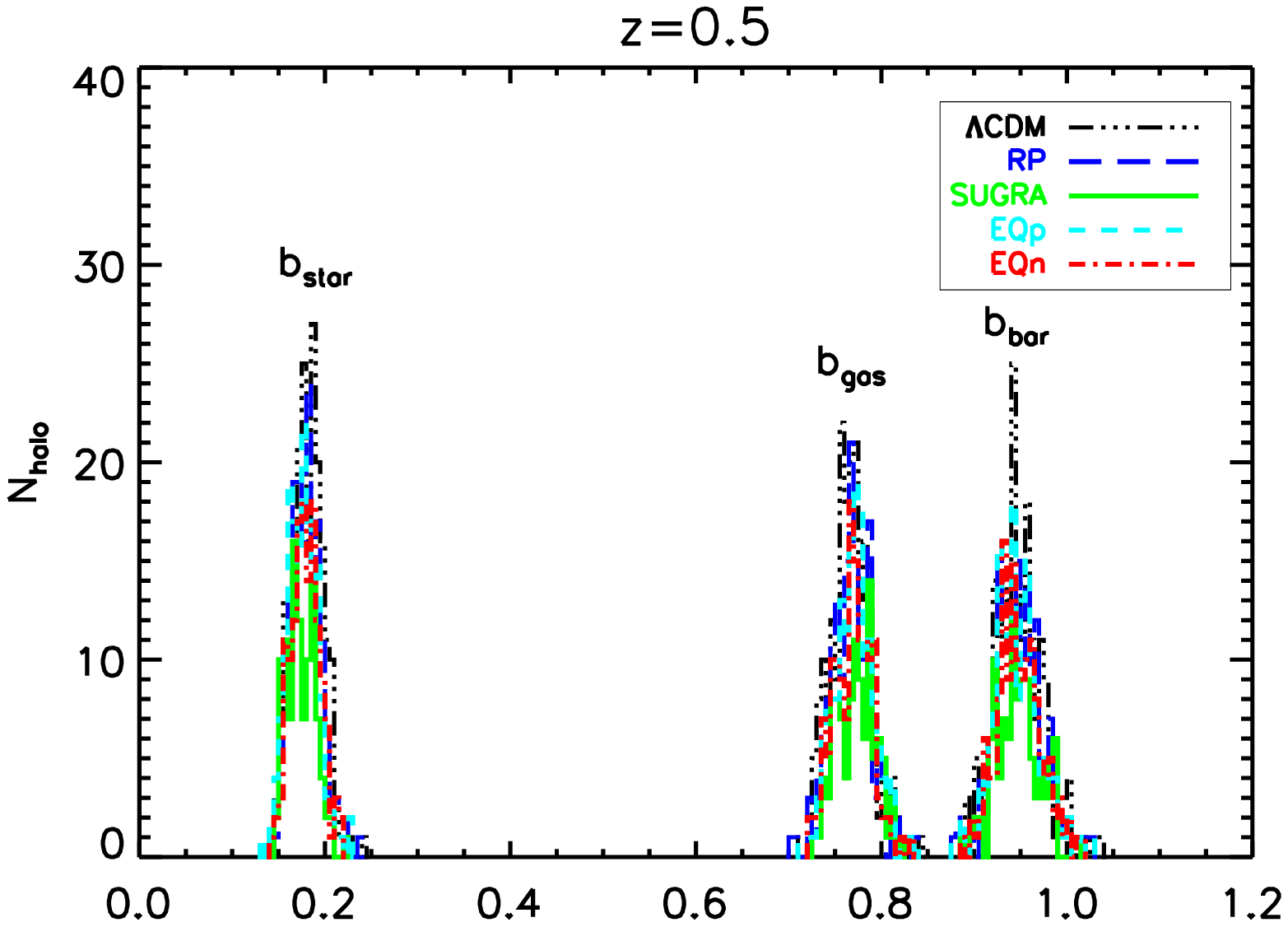,width=0.50\textwidth}
}
\hbox{
 \epsfig{figure=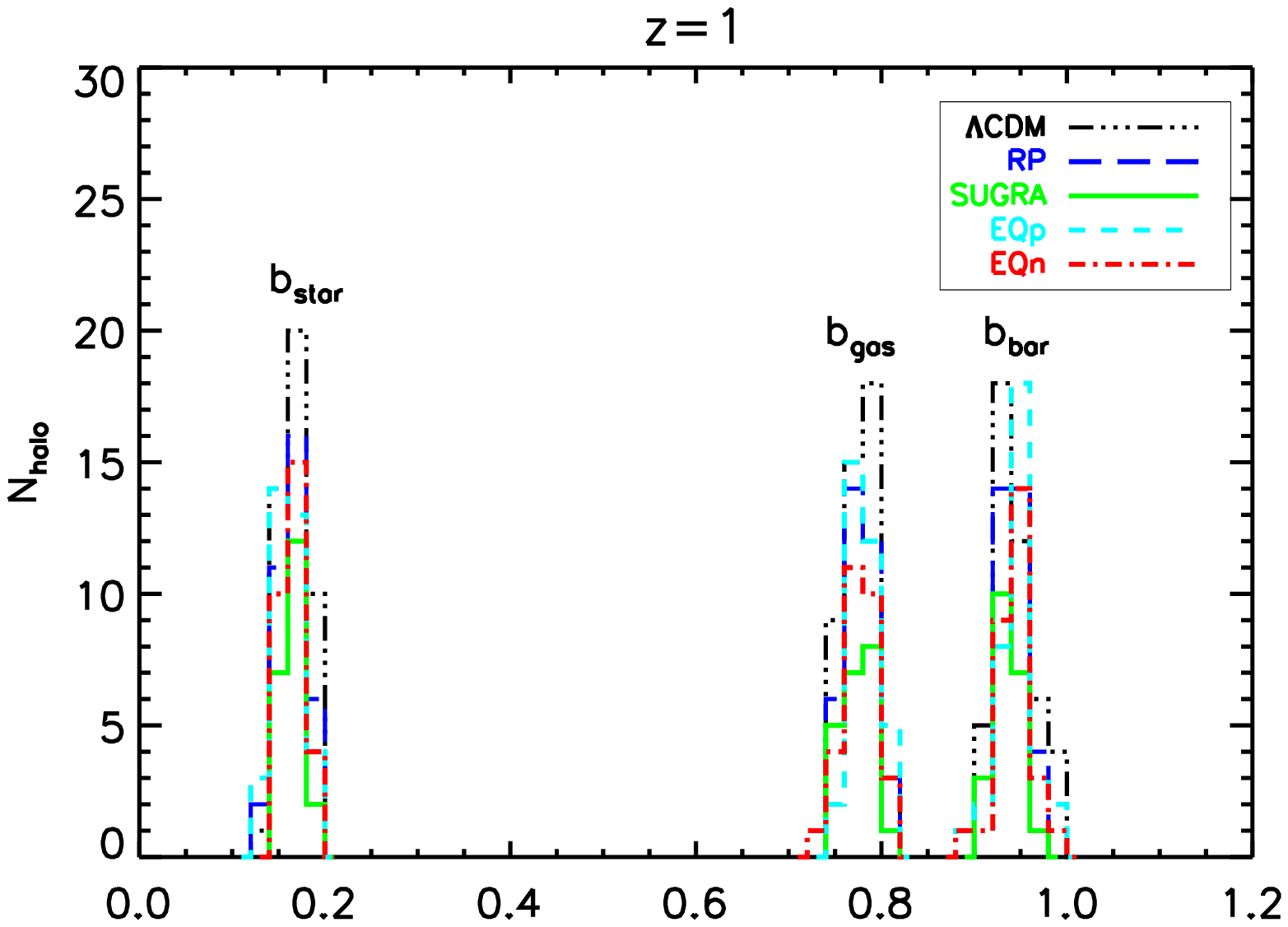,width=0.50\textwidth}
}
\caption{Distribution of $b_{bar}$, $b_{gas}$, and $b_{star}$ evaluated at $R_{200}$ for the $\Lambda$CDM (black), RP (blue), SUGRA (green), EQp (cyan), and EQn (red) cosmologies at $z=0$ (top panel), $z=0.5$ (middle panel), and $z=1$ (bottom panel).}
\label{histograms}
\end{figure}

In Fig. \ref{histograms} we plot the distribution of $b_{bar}$, $b_{gas}$ and $b_{star}$ evaluated at $R_{200}$ for the whole sample at $z=0$, $z=0.5$, and $z=1$ in order to check the spread of the values for the single objects.
We see that at $z=0$ there is a substantial overlapping among the different cosmologies, indicating that evolved objects have almost the same distribution whatever the underlying cosmological background is. The same is true looking at $z=0.5$ and $z=1$. We can note a decrease of $b_{gas}$ going from $z=1$ to $z=0$, not completely compensated by an increase of $b_{star}$. The net effect is a decrease of $b_{bar}$ going from $z=1$ to $z=0$.

In Table \ref{bbar} we summarize the mean value of $b_{star}$, $b_{gas}$, and $b_{bar}$ evaluated at $R_{2500}$, $R_{500}$, and $R_{200}$ for all the objects in the different cosmological models considered, at $z=0$, $z=0.5$, and $z=1$. We see that, on the one hand, for any cosmological model, at any redshift, $b_{star}$ is a decreasing function of radius, going from $R_{2500}$ to $R_{200}$. On the other hand, $b_{gas}$ is an increasing function of radius. As a whole, $b_{bar}$ is slightly decreasing with radius. Fixing the radius, either $R_{2500}$, $R_{500}$, or $R_{200}$, $b_{star}$ increases going from $z=1$ to $z=0$, while $b_{gas}$ decreases. As we already noted from Fig. \ref{histograms}, $b_{bar}$ is slightly decreasing going from $z=1$ to $z=0$. These trends are general, and they hold for all the cosmological models considered.

\begin{figure*}
\hbox{
 \epsfig{figure=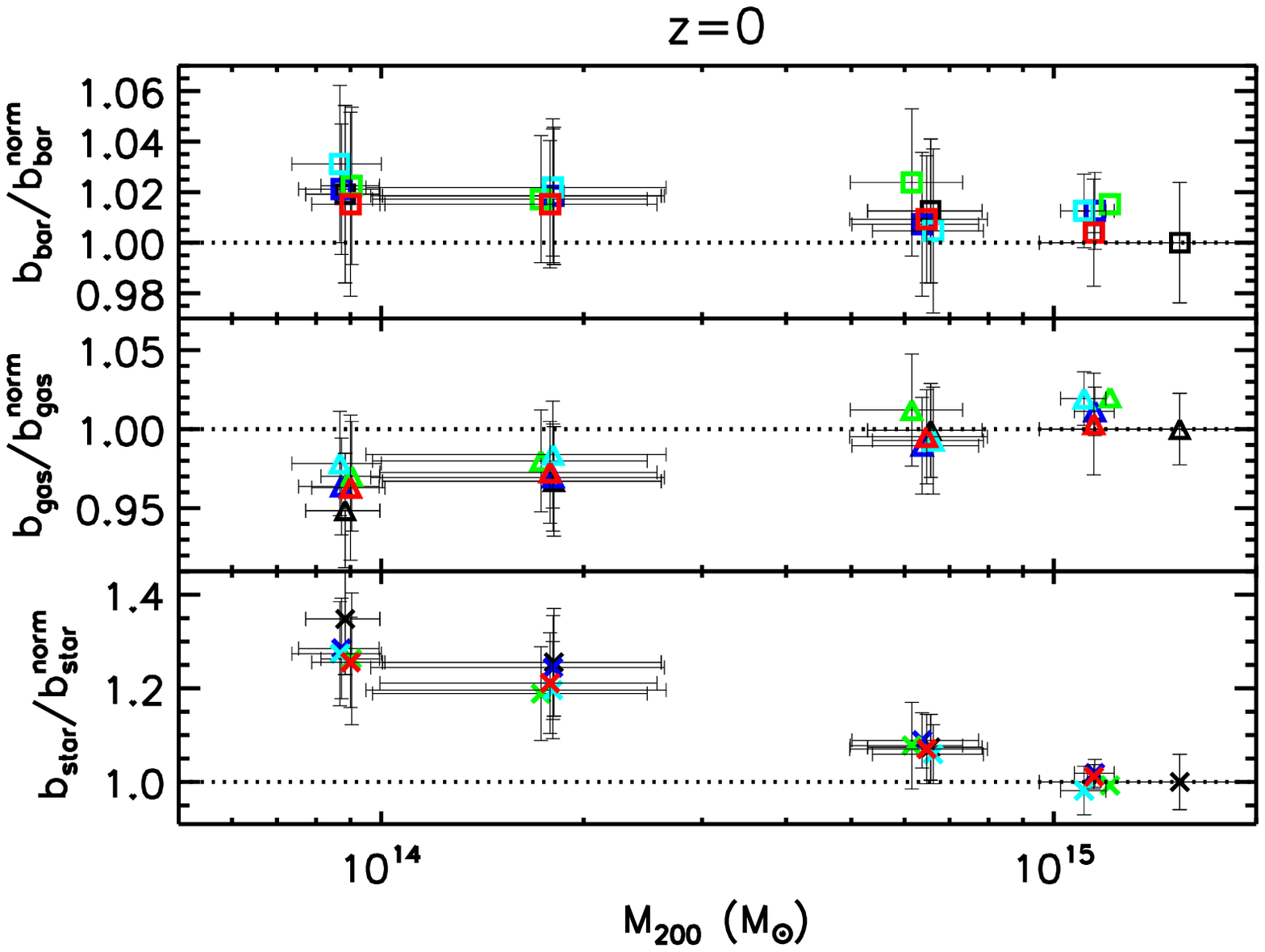,width=0.50\textwidth}
 \epsfig{figure=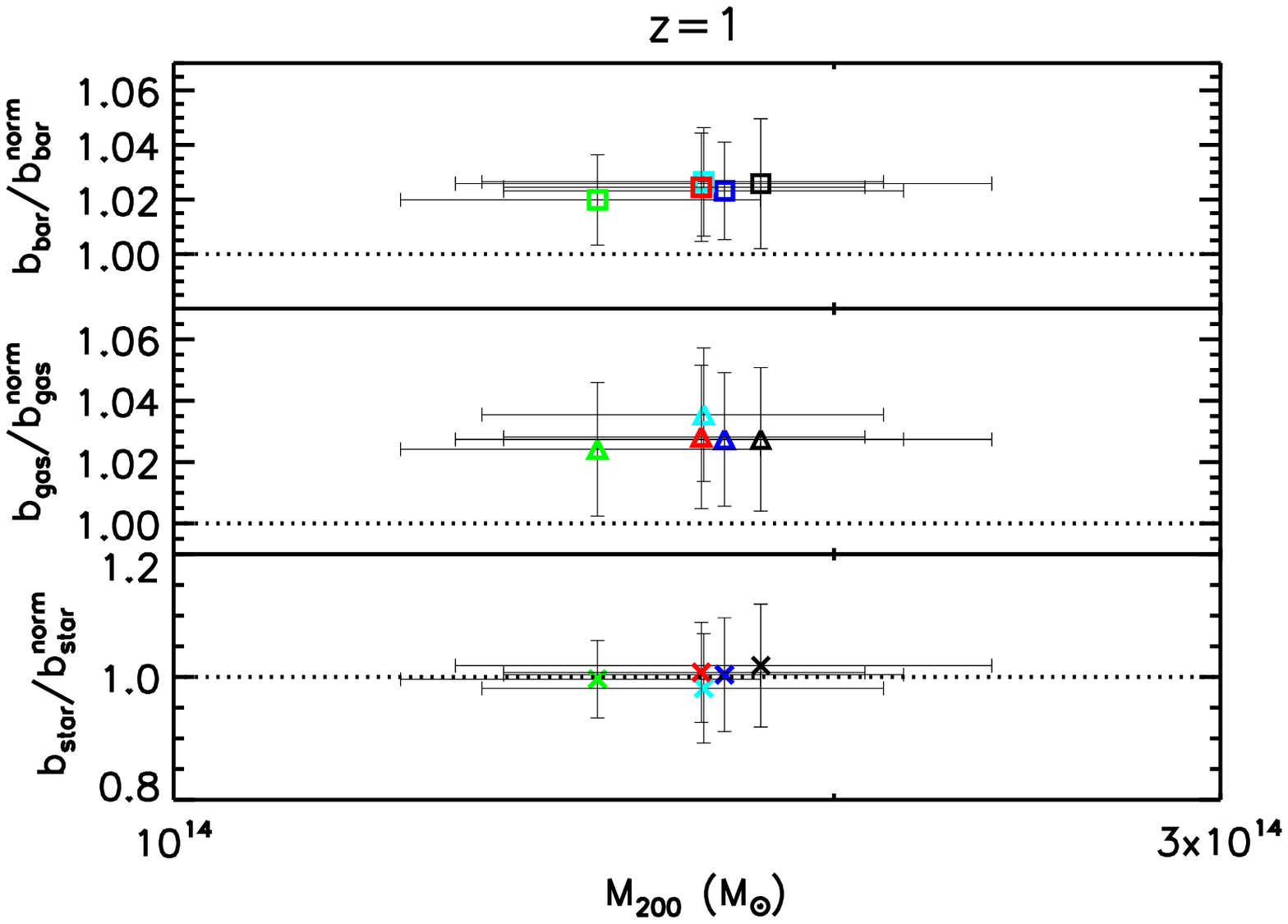,width=0.50\textwidth}
}
\caption{The evolution of stellar, gas and baryon depletion parameter evaluated at $R_{200}$ with mass at $z=0$ (left panel) and $z=1$ (right panel) for the $\Lambda$CDM (black), RP (blue), SUGRA (green), EQp (cyan), and EQn (red) cosmologies. Crosses, triangles, and squares indicate $b_{star}$, $b_{gas}$, and $b_{bar}$ respectively. The depletion parameters are expressed in units of $b^{norm}_{bar}$, $b^{norm}_{gas}$ and $b^{norm}_{star}$, the mean values for clusters with $M_{200}\ge10^{15}\rm{M_{\odot}}$ in the $\Lambda$CDM cosmology at $z=0$. Error bars are r.m.s. of the mean distribution.}
\label{mass_dependence}
\end{figure*}

\begin{table*}
\caption{Mean values of $b_{star}$, $b_{gas}$, and $b_{bar}$ evaluated at $R_{2500}$, $R_{500}$, and $R_{200}$ for all the objects in the different cosmological models considered, at $z=0$, $z=0.5$, and $z=1$. N indicates the number of objects in the given model at the given redshift. Numbers in brackets are $1\sigma$ errors on the mean.}
\label{}
\centering
\begin{tabular}{c c c c c c c c c c c c}
\hline
Model & $z$ & N & $b_{star2500}$ & $b_{gas2500}$ & $b_{bar2500}$ & $b_{star500}$ & $b_{gas500}$ & $b_{bar500}$ & $b_{star200}$ & $b_{gas200}$ & $b_{bar200}$ \\
\hline
$\Lambda$CDM & 0 & 563 & 0.508 & 0.535 & 1.043 & \ 0.269 & 0.680 & 0.948 & \ 0.207 & 0.731 & 0.937 \\
& & & (0.072) & (0.067) & (0.065) & \ (0.032) & (0.037) & (0.036) & \ (0.021) & (0.027) & (0.026) \\
& 0.5 & 202 & 0.461 & 0.578 & 1.039 & \ 0.236 & 0.724 & 0.961 & \ 0.182 & 0.767 & 0.949 \\
& & & (0.063) & (0.055) & (0.065) & \ (0.025) & (0.028) & (0.032) & \ (0.017) & (0.022) & (0.027) \\
& 1 & 45 & 0.454 & 0.624 & 1.078 & \ 0.222 & 0.749 & 0.971 & \ 0.168 & 0.777 & 0.944 \\
& & & (0.082) & (0.063) & (0.068) & \ (0.029) & (0.027) & (0.035) & \ (0.016) & (0.018) & (0.022) \\
\hline
RP & 0 & 484 & 0.498 & 0.541 & 1.039 & \ 0.263 & 0.683 & 0.946 & \ 0.204 & 0.733 & 0.937 \\
& & & (0.069) & (0.066) & (0.063) & \ (0.029) & (0.037) & (0.036) & \ (0.019) & (0.026) & (0.024) \\
& 0.5 & 164 & 0.452 & 0.589 & 1.041 & \ 0.234 & 0.726 & 0.960 & \ 0.181 & 0.767 & 0.948 \\
& & & (0.061) & (0.059) & (0.069) & \ (0.025) & (0.031) & (0.032) & \ (0.016) & (0.022) & (0.026) \\
& 1 & 35 & 0.429 & 0.633 & 1.063 & \ 0.218 & 0.754 & 0.972 & \ 0.165 & 0.777 & 0.942 \\
& & & (0.070) & (0.062) & (0.073) & \ (0.026) & (0.025) & (0.031) & \ (0.015) & (0.017) & (0.016) \\
\hline
SUGRA & 0 & 352 & 0.520 & 0.549 & 1.069 & \ 0.260 & 0.693 & 0.953 & \ 0.196 & 0.740 & 0.937 \\
& & & (0.076) & (0.066) & (0.066) & \ (0.029) & (0.036) & (0.035) & \ (0.018) & (0.025) & (0.024) \\
& 0.5 & 105 & 0.442 & 0.602 & 1.044 & \ 0.226 & 0.736 & 0.962 & \ 0.174 & 0.773 & 0.947 \\
& & & (0.066) & (0.054) & (0.064) & \ (0.025) & (0.028) & (0.029) & \ (0.015) & (0.021) & (0.024) \\
& 1 & 21 & 0.435 & 0.628 & 1.063 & \ 0.215 & 0.750 & 0.964 & \ 0.164 & 0.774 & 0.939 \\
& & & (0.063) & (0.069) & (0.089) & \ (0.021) & (0.027) & (0.040) & \ (0.010) & (0.016) & (0.015) \\
\hline
EQp & 0 & 476 & 0.515 & 0.554 & 1.069 & \ 0.258 & 0.689 & 0.947 & \ 0.197 & 0.744 & 0.941 \\
& & & (0.078) & (0.068) & (0.075) & \ (0.030) & (0.037) & (0.036) & \ (0.018) & (0.026) & (0.026) \\
& 0.5 & 162 & 0.444 & 0.597 & 1.042 & \ 0.228 & 0.731 & 0.959 & \ 0.176 & 0.770 & 0.946 \\
& & & (0.061) & (0.058) & (0.61) & \ (0.024) & (0.031) & (0.033) & \ (0.016) & (0.022) & (0.026) \\
& 1 & 34 & 0.429 & 0.625 & 1.054 & \ 0.213 & 0.760 & 0.973 & \ 0.162 & 0.783 & 0.944 \\
& & & (0.072) & (0.064) & (0.078) & \ (0.026) & (0.026) & (0.031) & \ (0.015) & (0.016) & (0.018) \\
\hline
EQn & 0 & 431 & 0.508 & 0.545 & 1.053 & \ 0.258 & 0.689 & 0.947 & \ 0.199 & 0.736 & 0.934 \\
& & & (0.070) & (0.058) & (0.070) & \ (0.028) & (0.034) & (0.034) & \ (0.018) & (0.025) & (0.024) \\
& 0.5 & 140 & 0.452 & 0.599 & 1.052 & \ 0.234 & 0.729 & 0.963 & \ 0.179 & 0.769 & 0.948 \\
& & & (0.063) & (0.063) & (0.054) & \ (0.024) & (0.028) & (0.032) & \ (0.016) & (0.020) & (0.024) \\
& 1 & 29 & 0.427 & 0.626 & 1.053 & \ 0.216 & 0.753 & 0.969 & \ 0.166 & 0.777 & 0.943 \\
& & & (0.068) & (0.065) & (0.071) & \ (0.026) & (0.026) & (0.031) & \ (0.013) & (0.017) & (0.018) \\
\hline
\end{tabular}
\label{bbar}
\end{table*}

In the left panel of Fig. \ref{mass_dependence} we plot, for each cosmology, the ratio between the mean values of $b_{bar}$, $b_{gas}$ and $b_{star}$ evaluated at $R_{200}$ in four different mass ranges at $z=0$ and the mean value of $b^{norm}_{bar}$, $b^{norm}_{gas}$ and $b^{norm}_{star}$ for clusters having $M_{200}\ge10^{15}\rm{M_{\odot}}$ in the $\Lambda$CDM cosmology at $z=0$ ({\it{i.e.}} $0.921$, $0.757$ and $0.165$ respectively). We have considered four mass ranges: $M_{200}<10^{14}{\rm{M_{\odot}}}$, $10^{14}{\rm{M_{\odot}}} \le M_{200}< 5 \times 10^{14}{\rm{M_{\odot}}}$, $5 \times 10^{14}{\rm{M_{\odot}}}\le M_{200}<10^{15}{\rm{M_{\odot}}}$, and $M_{200}\ge10^{15}\rm{M_{\odot}}$. We have evaluated the quantities at $R_{200}$ instead of $R_{500}$ as in Sect. \ref{LT} because this radius is representative of the cluster as a whole, including the external regions, and indeed we want to check whether, in different cosmologies, these objects are a fair representation of the underlying background. 
The first thing we notice is that, in every mass bin, the values of $b_{bar}$, $b_{gas}$ and $b_{star}$ are similar, within error bars, among the different cosmologies. So we can refer to a single cosmology ({\it{e.g.}} $\Lambda$CDM) in order to study the mass dependence of these quantities. We see that $b_{bar}$ is almost constant, independently of mass. On the one hand, $b_{gas}$ shows a slight positive trend, of the order of $5$\%, going from low-mass to high-mass systems, but still compatible with a constant value within the error bars. On the other hand, $b_{star}$ shows a decrease up to $30$\% going from low-mass to high-mass clusters, not compatible with a constant value.
In the right panel of Fig. \ref{mass_dependence} we plot, for each cosmology, the ratio between $b_{bar}$, $b_{gas}$ and $b_{star}$ evaluated at $R_{200}$ at $z=1$ and the mean values for $\Lambda$CDM at $z=0$ already used in the left panel. In this case we do not consider different mass ranges, since at this redshift the cluster abundance starts to be low and all the objects have $10^{14}{\rm{M_{\odot}}} \le M_{200}< 5 \times 10^{14}{\rm{M_{\odot}}}$. Again, the cosmologies are completely equivalent within error bars.

Here we stress again that our simulations do not follow AGN feedback. It is known from literature \citep[{\it{e.g}}.][]{2008ApJ...687L..53P} that the effect of this feedback is mass dependent, leading to a lowering in the baryon fraction in groups and low-mass clusters, without affecting significantly high-mass clusters.

We find in general a constant baryon fraction with respect to the mass. Some authors \citep[{\it{e.g}}.][]{2009ApJ...703..982G} claim that in observed objects the total baryon fraction shows an increase with increasing mass. This difference with respect to our results could be due to the fact that some relevant physical processes are still not included in our cosmological simulations. Such processes may be able to affect the global properties of low-mass systems without changing the high-mass clusters. Not including them in the simulations does not permit to us to fully compare our results with observations. In particular, we note an overabundance of stars (which obviously influences the total baryon fraction) in low-mass objects.

Combining the right and left panels of Fig. \ref{mass_dependence}, we can study the evolution with redshift of $b_{bar}$, $b_{gas}$ and $b_{star}$. Since the differences among various cosmologies at the same redshift are quite small, we rely on our reference $\Lambda$CDM model for the analysis of redshift evolution. For clusters with $10^{14}{\rm{M_{\odot}}} \le M_{200}< 5 \times 10^{14}{\rm{M_{\odot}}}$, the mean value of $b_{bar}$ is almost constant, with a slight decrease of about $2$\%, going from $z=1$ to $z=0$. In particular, $b_{gas}$ decreases of less than $10$\%, while the increase of $b_{star}$ is of the order of $25$\%. A decrease of the baryon fraction with decreasing redshift was already found in other simulations \citep[see {\it{e.g.}}][]{2006MNRAS.365.1021E}, and a possible explanation is that at high redshift the radius at which the baryons accrete is smaller than at low redshift, and so a greater number of baryons can fall in the cluster potential well.

\begin{figure}
\hbox{
 \epsfig{figure=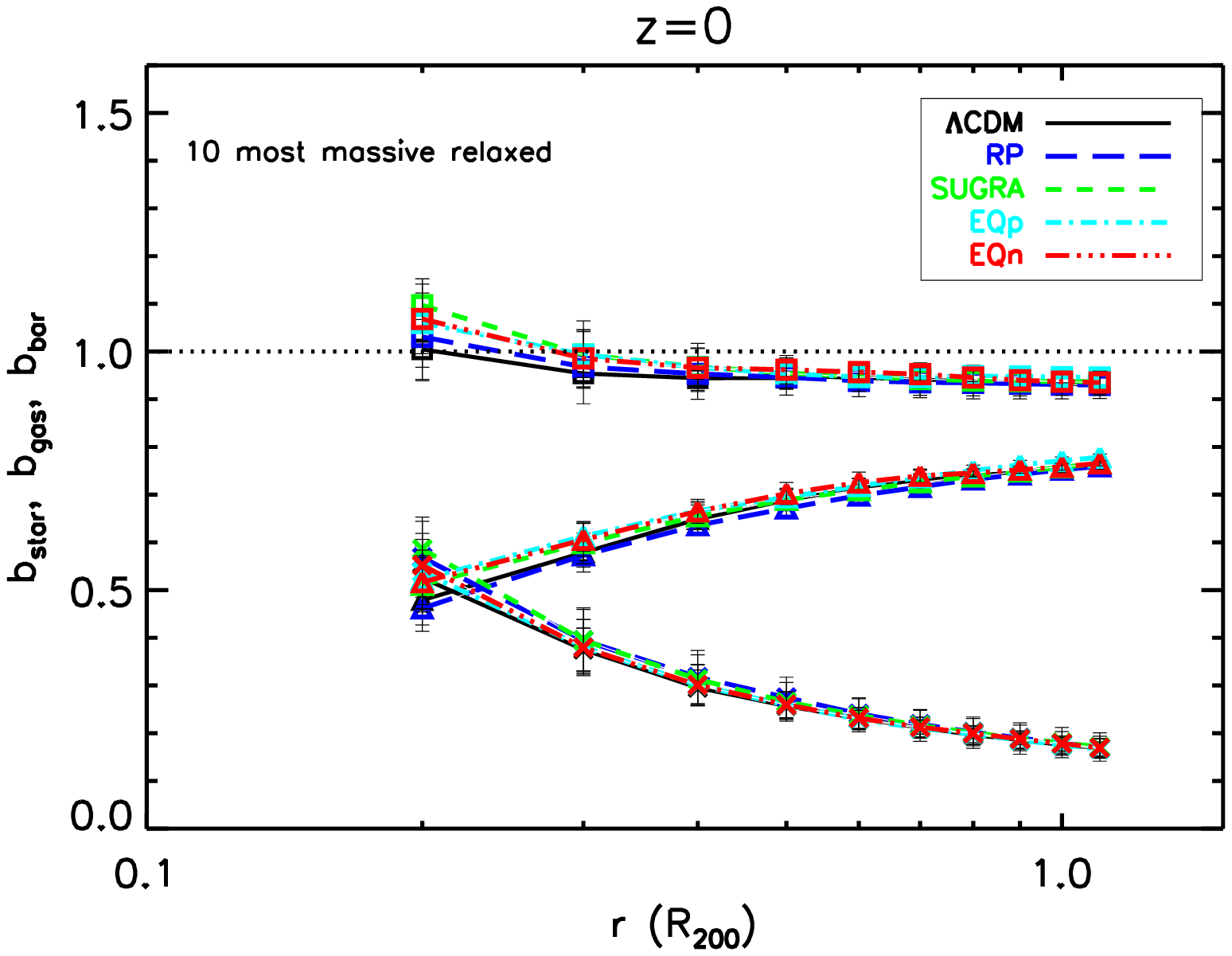,width=0.50\textwidth}
}
\hbox{ 
 \epsfig{figure=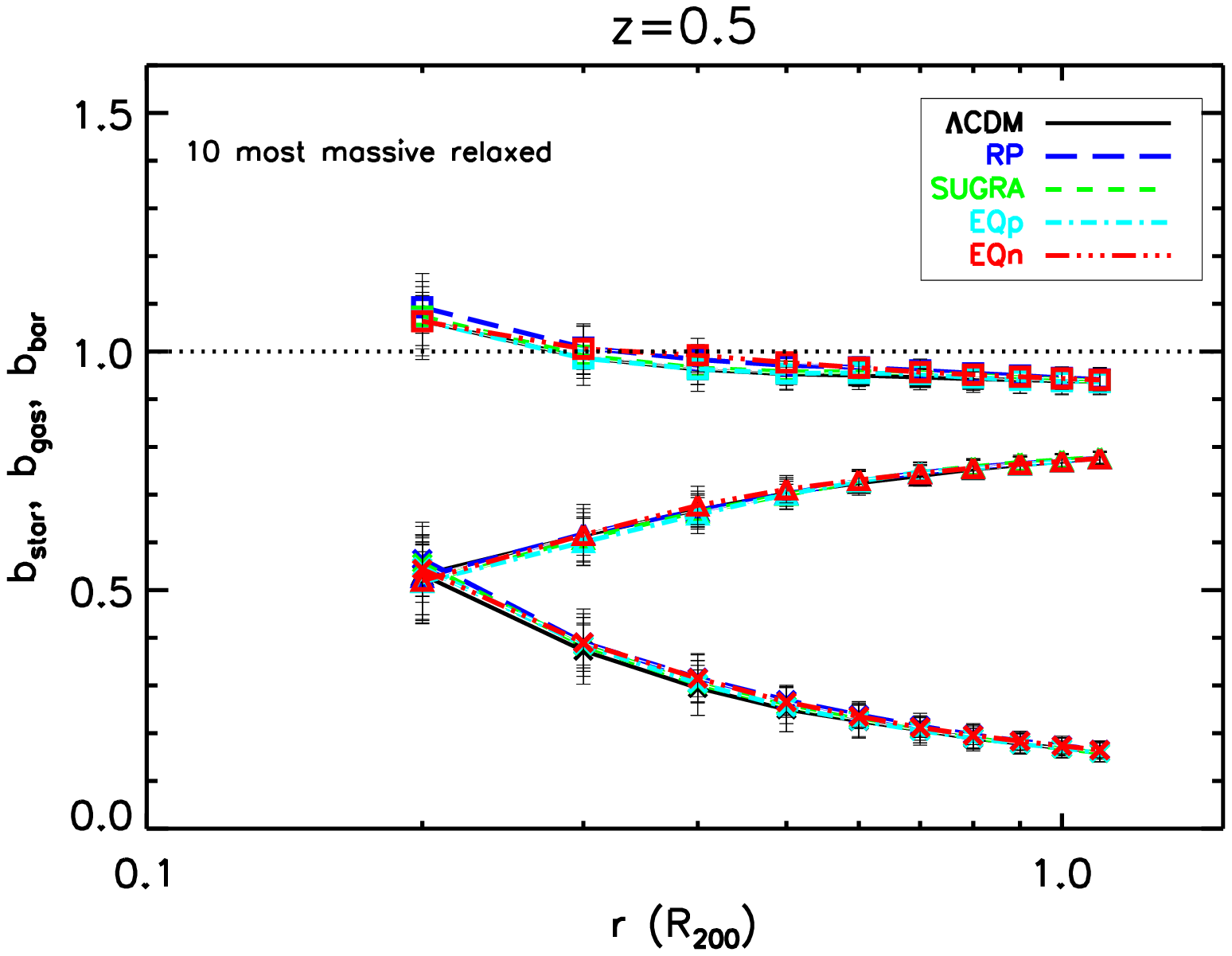,width=0.50\textwidth}
}
\hbox{  
 \epsfig{figure=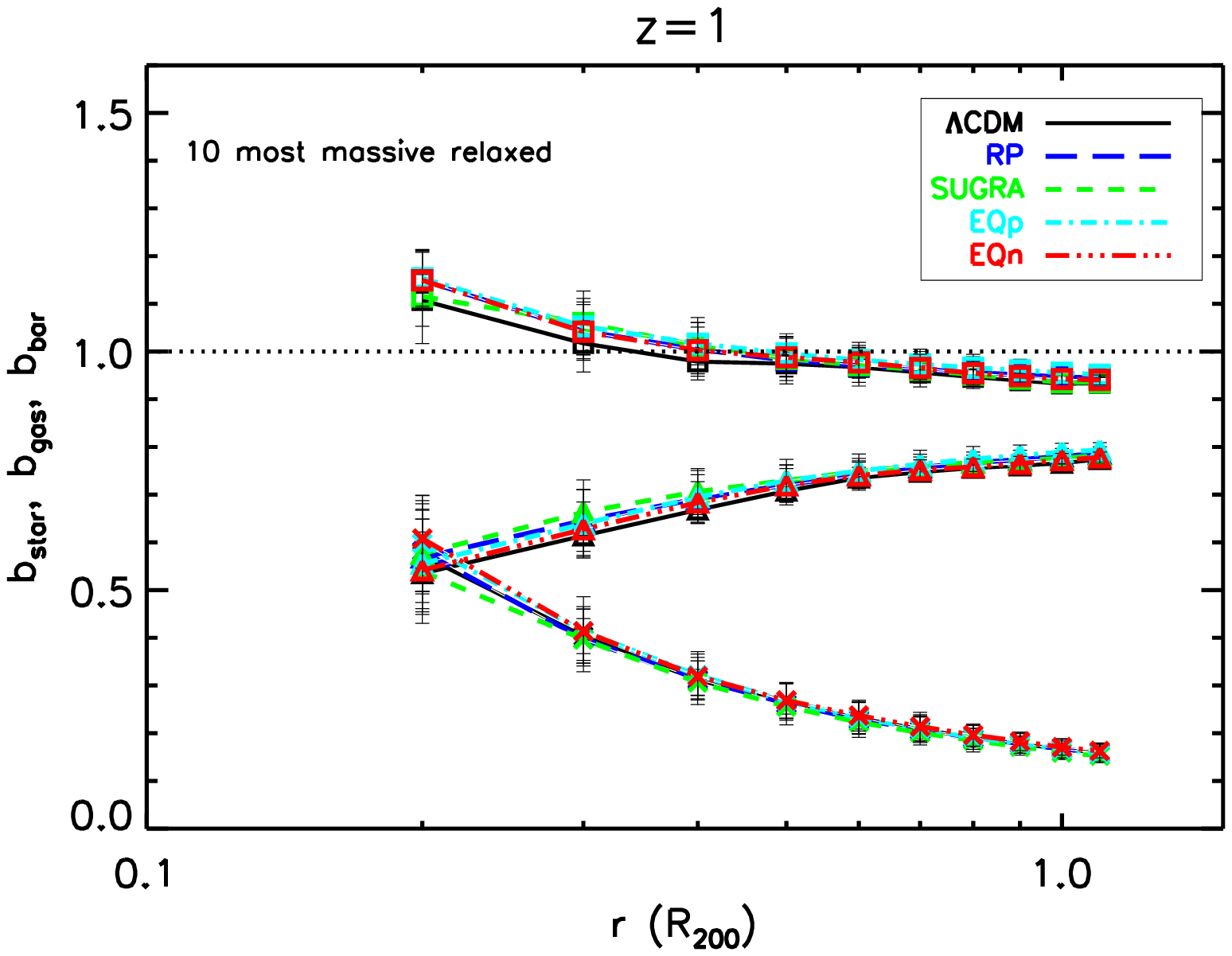,width=0.50\textwidth}
}
\caption{The evolution of stellar, gas and baryon depletion parameter with radius at $z=0$ (top panel), $z=0.5$ (middle panel), and $z=1$ (bottom panel) for an object obtained by stacking the ten most massive relaxed clusters in the $\Lambda$CDM (black), RP (blue), SUGRA (green), EQp (cyan), and EQn (red) cosmologies. Crosses, triangles, and squares indicate $b_{star}$, $b_{gas}$, and $b_{bar}$ respectively. The depletion parameters are expressed in units of the cosmic baryon fraction $\Omega_{0b}/\Omega_{0m}=0.164$. Error bars are r.m.s. of the mean distribution.} 
\label{radius_dependence}
\end{figure}

Finally, we study the star, gas, and baryon depletion parameters as a function of the distance from the centre of the cluster, defined as the position of the most bound particle. For each cosmology we select, as in Sect. \ref{LT}, the ten most massive (in $M_{200}$) relaxed halos and generate the radial profile of the object obtained by stacking them. We do this at $z=0$, $z=0.5$ and $z=1$. We recall that, at $z=1$, for SUGRA we only have six objects. The resulting profiles, expressed in units of the cosmic baryon fraction $\Omega_{0b}/\Omega_{0m}=0.164$, are shown in Fig. \ref{radius_dependence}. At $z=0$, in the outer regions near $R_{200}$, the five cosmologies are completely equivalent, with $b_{bar}$ having almost the cosmological value, while looking toward the centre some differences can be seen. This fact means that, as a whole, evolved relaxed clusters contain the same amount of baryonic matter, independently of the underlying cosmological model, but that the matter can be redistributed inside them according to their formation history. This fact is confirmed by looking at $z=0.5$ and in particular at $z=1$, where the differences among the models are clear even in the outer regions, indicating a sort of self-regulating mechanism that leads to the same objects at $z=0$ even if they can be very different at higher redshifts. Again, the same features appear both in the mean values of the whole sample and in more relaxed and massive objects, indicating that the latter are a fair representation of the clusters in a given cosmological model.

As a general rule for the radial profiles, it is confirmed the well known relative trend of the radial profile of gas and stars components, being the former increasing with radius and the latter decreasing, giving a total baryon fraction almost constant (but slightly decreasing) beyond $0.5 R_{200}$. Then we note that the total baryon fraction at $z=1$ is higher compared to $z=0$, in particular in the inner regions of clusters. The effect is mainly due to a higher star fraction in the inner regions of clusters at $z=1$. Another quite evident feature is that the radius at which the gas starts to dominate over the stars increases with increasing redshift. The explanation is that, as we have just seen, the gas fraction profile is almost independent of redshift, while the star fraction at a given radius increases with redshift, and so at higher redshift it remains the dominant baryonic component also at larger radii.

\section{Conclusions}  \label{conclusions}

In this paper we have analysed the general properties of a sample of galaxy clusters extracted from hydrodynamical simulations of different cosmological models with dynamical dark energy. We simulate a cosmological box of size $(300 \ {\rm{Mpc}} \ h^{-1})^{3}$, resolved with $(768)^{3}$ dark matter particles and the same amount of gas particles. The reference cosmology is a concordance $\Lambda$CDM model normalized to WMAP3 data. The others are two quintessence models, one with a RP and the other with a SUGRA potential, and two extended quintessence models, with a positive and a negative coupling between quintessence and gravity, indicated as EQp and EQn, respectively. Since all models are normalized to CMB data, they have different $\sigma_{8}$, and thus different structure formation histories. We focus on various properties of the considered objects, in particular the mass function, the X-ray $L-T$ relation, the X-ray luminosity and temperature functions (XLF and XTF respectively) and finally the baryonic fraction in terms of the depletion parameters $b_{star}$, $b_{gas}$ and $b_{bar}$ defined in equation (\ref{depletion}). We select and study objects at three different redshifts, $z=0$, $z=0.5$, and $z=1$, with $M_{200m} \ge 1.42 \times 10^{14}{\rm{M_{\odot}}}$. We also define a criterion to distinguish between relaxed and unrelaxed clusters. From our analysis we draw the following conclusions.

\begin{itemize}
\item Mass function: at $z=0$ the total mass function evaluated at $R_{200m}$ shows different behaviours in the different cosmologies, in particular in the normalization. The $\Lambda$CDM model tends to form more clusters of a given mass with compared to the other cosmologies; SUGRA is the cosmology with the smallest abundance, while RP, EQp and EQn lie in between, with RP and EQp being the closest to $\Lambda$CDM. This fact directly reflects the redshift evolution of the equation of state parameter $w$ and of the growth factor, given the different assumed value of $\sigma_{8}$ in the various models. Actually, for extended quintessence models, a positive value of the coupling leads to an higher linear density contrast, and vice versa for a negative coupling. This would imply a higher mass function for models with negative coupling ({\it{i.e.}} EQn) than for models with positive coupling ({\it{i.e.}} EQp), keeping fixed all remaining parameters. In our case, this effect is somehow mitigated by the different $\sigma_{8}$ used. This trend is preserved also at $z=0.5$ and $z=1$. The differences among the models are more pronounced in the high-mass tail of the distribution. This is expected, because very massive objects form later and are representative of the different structure formation time scale of the considered cosmologies. We note here that our results are different from what has been found in the case of coupling with dark matter \citep{2011MNRAS.412L...1B}, where there is an enhancement in the number counts of massive objects.
\item $L-T$ relation: we compare the $L-T$ relation of our simulated objects in the $\Lambda$CDM reference models with a collection of observed objects \citep{2009A&A...498..361P}. Despite the differences in the slope of the relation in the two cases ($1.81$ for our simulated objects vs $2.53 \pm 0.16$ for their observed ones), we find that there is a good agreement in the high-temperature high-luminosity region where X-ray selected observed objects are found. The discrepancy in the low-temperature low-luminosity region is not worrying, because low-mass systems are globally more affected by physical mechanisms not yet fully understood and reproduced \citep[{\it{e.g}}.][]{2004MNRAS.348.1078B}, acting in the core. We also study the evolution with redshift of the $L-T$ relation for the ten most massive relaxed objects in each cosmology, both keeping and cutting the core. We find that cutting the core results in both a lower mean luminosity and lower mean emission-weighted temperature. In general, both the mean luminosity and temperature increase with decreasing redshift, independently of the cosmological model, because they trace the hierarchical growth of structures.
\item X-ray observable functions: the relative behaviour observed in the mass functions is also qualitatively reproduced by the XLFs and XTFs evaluated in the [$0.5 - 2$] keV band in the [$0.15 - 1$]$R_{500}$ region, with few exceptions. In particular, in the range of luminosity around $0.5 \times 10^{43} \rm{erg \ s}^{-1}$ RP tends to form $10$\% more clusters than $\Lambda$CDM. We also check the X-ray $M_{gas500}$ and $Y_{X500}$ functions as proxies for the mass function. We conclude that all the X-ray observable functions are more or less equivalent, with $T_{ew}$ and $M_{gas500}$ being slightly more stable than $L_{X}$ and $Y_{X500}$, in tracing the mass function and thus disentangle the growth of structures among
different dark energy models. For each dark energy model we evaluate the volumes that a cluster survey must cover in order to be able to distinguish it from the concordance $\Lambda$CDM model, using the mass function, the XLF, and the XTF.
\item Baryon fraction: the analysis of the $b_{star}$, $b_{gas}$, and $b_{bar}$ dependence on mass, redshift and distance from the cluster centre shows that there is no significant difference among the five cosmologies considered, if we limit ourselves to the values at $R_{200}$ and at $z=0$. Therefore, at these conditions, $b_{bar}$ (and so the baryon fraction $f_{bar}$) can be safely used as a cosmological proxy to derive the value of other cosmological parameters. In addition, we do not find any clear positive trend of the total baryon fraction with mass, while we see a positive trend (of the order of $5$\%) of the gas fraction and a negative trend (of the order of $30\%$) of the star fraction going from low-mass to high-mass systems. Considering observations of real objects, in spite of finding the same trend for the gas and star fraction as we do, other authors \citep[{\it{e.g}}.][]{2009ApJ...703..982G} claim that the total baryon fraction is increasing with increasing mass. Actually, for all the cosmological models here considered, we find a slight decrease in the total baryon fraction with increasing mass. Still, we have to recall that, despite the hydrodynamical treatments in the simulations is based on sophisticated physical models, we do not include AGN feedback in our simulations. It is known from literature \cite[see {\it{e.g.}}][]{2008ApJ...687L..53P} that AGN feedback is mass dependent, in the sense that it globally affects more low-mass systems than high-mass systems. The net effect is the lowering of the total baryon fraction in low-mass objects while not affecting more massive clusters. Finally, we find a slight decrease (at most $5$\%) of the baryon fraction going from high to low redshift. A similar trend was already noted by \cite{2006MNRAS.365.1021E} and a possible explanation is that at high redshift the radius at which the baryons accrete is smaller than at low redshift, and so a greater number of baryons can fall in the cluster potential well.
\end{itemize}

In the end, we can conclude that in models with dynamical dark energy, the evolving cosmological background leads to different star formation rates and different formation histories of galaxy clusters, but the baryon physics is not affected in a relevant way. Indeed, evolved and relaxed clusters, if studied in regions sufficiently far from the centre, reveal to be very similar despite the different dark energy models considered. So, in conclusion, galaxy clusters can effectively be used as a probe to distinguish among different dark energy models.

\section*{Acknowledgments}

Computations have been performed at the ``Leibniz-Rechenzentrum''
with CPU time assigned to the Project ``h0073''.
We acknowledge financial contributions from contracts ASI I/016/07/0 COFIS,
ASI-INAF I/023/05/0, ASI-INAF I/088/06/0, ASI
`EUCLID-DUNE' I/064/08/0, PRIN MIUR 2009 ``Dark energy and cosmology with
large galaxy survey'', and PRIN INAF 2009 ``Towards an Italian network of computational cosmology''.
K.~D.~acknowledges the support of the DFG Priority Programme 1177
and additional support by the DFG Cluster of Excellence ``Origin
and Structure of the Universe''. We thank the anonymous referee for helping us improving the presentation of the results of our paper. 
We thank Matthias Bartelmann, Francesco Pace, Marco Baldi, Elena Rasia, Veronica Biffi, Stefano Borgani, and Cosimo Fedeli for useful discussions.

\end{document}